\documentclass[sigconf,screen, nonacm]{acmart}
\usepackage{threeparttable}
\usepackage[ruled,vlined]{algorithm2e}
\usepackage{enumitem}
\usepackage{subcaption}
\usepackage{booktabs}
\usepackage{multirow}
\usepackage{makecell} 
\usepackage{array} %

\AtBeginDocument{%
  }

\setcopyright{acmlicensed}
\copyrightyear{2018}
\acmYear{2018}
\acmDOI{XXXXXXX.XXXXXXX}

\acmConference[Recsys 2024]{Make sure to enter the correct
  conference title from your rights confirmation emai}{2024}{Bari, Italy}
\acmISBN{978-1-4503-XXXX-X/18/06}

\begin{document}

\title[Content-Agnostic Moderation]{Content-Agnostic Moderation \\for Stance-Neutral Recommendations}
\author{Nan Li}
\email{nan.li@ugent.be}
\affiliation{%
  \institution{Ghent University}
  \streetaddress{}
  \city{Ghent}
  \country{Belgium}
  \postcode{9000}
}

\author{Bo Kang}
\email{bo.kang@ugent.be}
\affiliation{%
  \institution{Ghent University}
  \streetaddress{}
  \city{Ghent}
  \country{Belgium}
  \postcode{9000}
}

\author{Tijl De Bie}
\email{tijl.debie@ugent.be}
\affiliation{%
  \institution{Ghent University}
  \streetaddress{}
  \city{Ghent}
  \country{Belgium}
  \postcode{9000}
}

\renewcommand{\shortauthors}{Li et al.}

\begin{abstract}
    Personalized recommendation systems often drive users towards more extreme content, exacerbating opinion polarization. While (content-aware) moderation has been proposed to mitigate these effects, such approaches risk curtailing the freedom of speech and of information. To address this concern, we propose and explore the feasibility of \emph{content-agnostic} moderation as an alternative approach for reducing polarization. Content-agnostic moderation does not rely on the actual content being moderated, arguably making it less prone to forms of censorship. We establish theoretically that content-agnostic moderation cannot be guaranteed to work in a fully generic setting. However, we show that it can often be effectively achieved in practice with plausible assumptions. We introduce two novel content-agnostic moderation methods that modify the recommendations from the content recommender to disperse user-item co-clusters without relying on content features.

    To evaluate the potential of content-agnostic moderation in controlled experiments, we built a simulation environment to analyze the closed-loop behavior of a system with a given set of users, recommendation system, and moderation approach. Through comprehensive experiments in this environment, we show that our proposed moderation methods significantly enhance stance neutrality and maintain high recommendation quality across various data scenarios%
. Our results indicate that achieving stance neutrality without direct content information is not only feasible but can also help in developing more balanced and informative recommendation systems without substantially degrading user engagement.
\end{abstract}

\keywords{}

\maketitle

\section{Introduction}
Personalized recommendation systems, which are particularly prevalent in news dissemination and social networking platforms, significantly influence public opinion dynamics and societal fragmentation. As they tend to present users with content that is similar to what they previously consumed, they can inadvertently affirm or reinforce users' stances on contentious topics, thus risking to exacerbate opinion polarization, skewing user perspectives and amplifying societal divides \cite{rossi2021closed,dandekar2013biased}. Conversely, recommending stance-diversified news can reduce selective exposure and mitigate polarization \cite{heitz2022benefits,knudsen2023modeling}.

These observations justify the development and use of content moderation strategies \cite{grimmelmann2015virtues} that filter or demote particular recommendations so as to mitigate the risk of polarization, for example by suppressing content on the extreme ends of the stance spectrum. That content moderation is regarded as an indispensable tool is evident from recent legislation, such as from the EU's Digital Services Act (DSA) \cite{eu2020digital} which may require social media platforms to moderate content that is assumed to pose so-called \emph{systemic risks}. A concern with such legislation is that it may curtail fundamental rights, such as the freedom of expression and of information \cite{turillazzi2023digital}. Indeed, the content moderation techniques used in practice, and envisioned by such laws, tend to selectively reduce exposure of particular content, which carries risks of real or perceived censorship.

Thus, our \emph{overall goal} is to investigate ways to reduce the risk of polarization without negatively impacting on engagement. We aim to avoid any measures that suppress or promote items based on content or other intrinsic properties, such as the information or stance it conveys. In contrast, the \emph{content-agnostic moderation} strategies we advocate utilize exclusively relational properties, such as user-item interactions (e.g., clicks and exposures).

We begin with investigating whether the use of content-unaware recommenders, which do not directly rely on the actual content, lead to reduced polarization associated with content features when compared to content-based recommenders. 
Secondly, we propose and investigate, both theoretically and empirically, the effectiveness of \emph{content-agnostic moderation techniques} for reducing polarization, i.e. moderation techniques that explicitly avoid using any intrinsic content, but leverage only relational properties.

Experimenting with various content recommender or moderation techniques is non-trivial, as it requires access to a large (social) media platform, and there are considerable ethical implications. To overcome these challenges in this and in future work, we developed a \emph{simulation environment} that allows one to empirically investigate the effect of various design choices of the recommendation system and the moderator, in the presence of various media biases and with respect to a range of user models.
\begin{figure}[hbt]
    \captionsetup{font=footnotesize}
    \centering
    \includegraphics[width=0.8\linewidth]{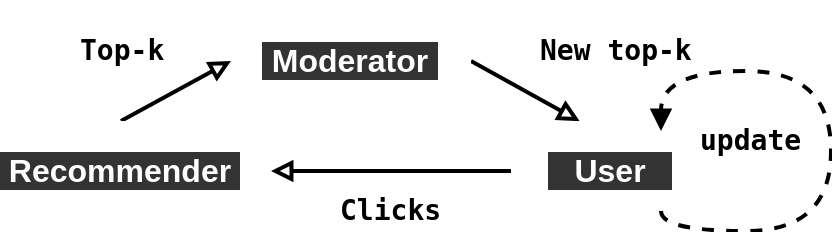}
    \caption{The feedback loop of the simulation framework.}
    \label{fig:feedback_loop}
\end{figure}

Using this simulation environment, we addressed three main research questions:
\begin{enumerate}[label={RQ\arabic*},topsep=5pt,itemsep=0pt,parsep=2pt,partopsep=5pt,align=parleft,left=0pt]
\item\label{rq1} To what extent can content-unaware recommenders (e.g., Matrix Factorization, Popularity-based recommendation) maintain stance neutrality? How do they compare with content-based models in neutrality and accuracy?
\item\label{rq2} Can content-agnostic moderation methods restore stance neutrality by post-processing top-$k$ recommendations, both theoretically (RQ2a) and empirically (RQ2b) under various simulation settings?
\item\label{rq3} What impacts does moderation have on user polarization, and how does that depend on the distribution of stances over the content that circulates in the media environment?
\end{enumerate}

Thus, our key contributions are:
\begin{enumerate}[topsep=5pt,itemsep=0pt,parsep=2pt,partopsep=5pt,align=parleft,left=0pt]
    \item We introduce an open-source \emph{simulation environment} and associated dataset, which enables one to empirically investigate the effect of various recommender system and content moderation design choices across a diversity of settings.
    \item We \emph{theoretically demonstrate} that while content-agnostic moderation cannot guarantee stance neutrality in generic settings without specific assumptions, it is feasible with favorable data or learnable proxy derived from relational properties.
    \item We \emph{empirically demonstrate} that content-agnostic moderation \emph{can} enhance stance neutrality, particularly in specific data configurations or when proxies for stance information are effectively derived from user-item interactions. %
    \item We introduce and evaluate two novel content-agnostic moderation methods: \emph{Random Dispersal} and \emph{Similarity-based Dispersal}. We empirically demonstrate that these methods can significantly improve stance neutrality while maintaining satisfactory CTR in a broad variety of conditions.
\end{enumerate}

We begin by reviewing related work in Sec.~\ref{sec:related_work}, followed by a theoretical analysis of content-agnostic moderation in Sec.~\ref{sec:theoretical_analysis}. We then describe our simulation framework in Sec.~\ref{sec:experiment}, including the data, recommendation models, moderators, and users. Next, we present the results of our experiments in Sec.~\ref{sec:results}, addressing our research questions. Finally, we conclude our work in Sec.~\ref{sec:conclusion}.

\section{Related work}\label{sec:related_work}
This research intersects disciplines such as computer science and social sciences, focusing on how news recommendation systems interact with user behavior, opinion dynamics, and polarization. 

The most directly related research explores \textbf{the interaction between news recommendation systems and user dynamics.}
Studies by Liu et al.~\cite{liu2021interaction} and Zhang et al.~\cite{zhang2023evolution} demonstrate how recommendation models can amplify extreme stances through feedback loops. Dandekar et al.~\cite{dandekar2013biased} and Spinelli et al.~\cite{spinelli2017closed} discuss how recommendations can polarize user opinions under certain conditions, with Jiang et al.~\cite{jiang2019degenerate} and Rossi et al.~\cite{rossi2021closed} further showing how personalized recommendations push opinions toward extremeness.

While this prior work primarily focuses on evaluating and auditing existing systems, our research is dedicated to the design and evaluation of content-agnostic moderation strategies. We specifically disentangle the effects of item exposure by recommenders from users' preferences, concentrating on moderating the former. Moreover, our work delves deeper into the effect of various media landscapes, providing a more comprehensive analysis than typically seen in simplistic models or limited simulations.

Research in digital media often addresses the phenomena of \textbf{filter bubbles and echo chambers}. Geschke et al.~\cite{geschke2019triple} introduced the concept of the triple-filter bubble, where individual, societal, and systemic factors all contribute. Bressan et al.~\cite{bressan2016limits} used a mathematical model to show market evolution under simplified popularity-based recommendation systems, leading to item share convergence. Chitra and Musco~\cite{chitra2020analyzing} examined how interventions aimed at reducing disagreements could actually increase user polarization in social networks. Additionally, Pansanella et al.~\cite{pansanella2023mass} highlighted that polarization and fragmentation effects are contingent on the population's open-mindedness and the balance of the information environment. Empirical studies by Nguyen et al.~\cite{nguyen2014exploring} and Bakshy et al.~\cite{bakshy2015exposure} demonstrated how recommendation algorithms can narrow the diversity of engagements and exposures, with Flaxman et al.~\cite{flaxman2016filter} observing potential increases in exposure to diverse views.

Our research contributes to these studies by focusing on a proactive, algorithmic approach to designing moderation mechanisms tailored for realistic recommendation scenarios. Unlike the observational or theoretical analyses common in computational social science, our work has flexibility in integrating different user models. Furthermore, we avoid the reliance on abstractions like filter bubbles and echo chambers, opting instead to precisely quantify the distribution of stances in both recommended and consumed items through formal definitions and metrics, recognizing that filter bubbles, echo chambers, and homogenization, though often conflated, are distinct concepts (further discussed in \cite{nguyen2020echo} and \cite{anwar2024filter}).

Specific to \textbf{news recommendation}, Patankar et al.~\cite{patankar2019bias} and Liu et al.~\cite{liu2023topic} address bias and topic awareness in recommendation. However, unlike our work, their methods rely heavily on content information.
Regarding \textbf{polarization}, Stray~\cite{stray2021designing} offers design guidelines for mitigating polarization in recommendation systems. However, these proposals remain conceptual. Badami et al.~\cite{badami2017detecting, badami2017peeking} focused on detecting polarizing items based on user ratings rather than content, exploring an uncommon interpretation of polarization related to user rating extremities. This is an enlightening perspective, yet distinct from ours in this paper.

Among the \textbf{post-processing or moderation techniques} of recommendation, approaches such as popularity suppression, item diversity improvement, and multi-sided fairness are somewhat related to content neutrality. Researchers like Seymen et al.~\cite{seymenunified2021}, Jambor and Wang~\cite{jambor2010optimizing}, and others have framed popularity suppression within a knapsack constraint optimization problem. Zhu et al.~\cite{zhu2021popularity} developed a system to elevate less exposed items by introducing popularity compensation scores. From a user-centric perspective, Abdollahpouri et al.~\cite{abdollahpouri2021user} calibrated recommendation scores to align with users' preferences across varying item popularity levels. Addressing individual user welfare, Jugovac et al.~\cite{jugovac2017efficient} and Patro et al.~\cite{patro2020fairrec} proposed algorithms, including a greedy RoundRobin procedure, to meet specific fairness metrics. However, none of these strategies directly target stance neutrality as we do in the current paper, and many depend on specific user or item profiles.

\textbf{Content moderation methods} are often categorized into ``hard moderation'' and ``soft moderation.'' Hard moderation techniques, reviewed in Gorwa et al.~\cite{gorwa2020algorithmic}, include algorithms based on matching~\cite{datar2004locality, davis2019open} or classification~\cite{schmidt2017survey}. Soft moderation methods, described by Narayanan et al.~\cite{narayanan2023understanding}, typically involve recommendations reduction~\cite{gillespie2022not} or content enrichment~\cite{chuai2023roll}, all necessitating content as input. For a comprehensive overview, see Gongane et al.~\cite{gongane2022detection}. 
In contrast with these approaches, our research uniquely pursues the challenging goal of content-agnostic moderation, stepping beyond conventional content dependency.

\section{Theoretical Analysis}\label{sec:theoretical_analysis}
This section examines the limitations of post-processing moderation strategies for achieving a specific distribution of recommended item stances without content or stance information. We show that such moderation cannot be guaranteed in generic settings (Section~\ref{sec:theoretical_impossibility}). Additionally, we explore the conditions under which achieving stance neutrality is empirically feasible through content-agnostic moderation (Section~\ref{sec:empirical_possibility}). %

\subsection{Non-Determinacy Theorem on Content-Agnostic Moderation}\label{sec:theoretical_impossibility}

\begin{definition}[Problem setup]
    Let \(\mathcal{U}\) denote a set of users and \(\mathcal{I}\) a set of items within a recommendation system. Each item \(i \in \mathcal{I}\) possesses intrinsic properties \(\mathcal{P}_i^{\text{int}}\), such as content and stance \(s_i \in \mathcal{S}\) (where \(\mathcal{S}\) is a finite set of stances and \(s_i\) is typically inferred from the content), and relational properties \(\mathcal{P}_i^{\text{rel}}\), including click counts by users and exposure frequencies.
    \footnote{Similarly, each user \(u \in \mathcal{U}\) is characterized by intrinsic properties \(\mathcal{P}_u^{\text{int}}\) (e.g., demographics and latent preferences towards stances) and relational properties \(\mathcal{P}_u^{\text{rel}}\) (e.g., interaction frequencies with items).}
\end{definition}

\begin{definition}[Content-Agnostic Moderation Function]
    Define \(\pi_u\) as an ordered list of items recommended to user \(u\), and let \(\pi = \{\pi_u \mid u \in \mathcal{U}\}\) represent the set of these lists, forming the overall recommendation configuration for users \(\mathcal{U}\) and items \(\mathcal{I}\). The function \(f: \Pi \rightarrow \Pi\) is to modify \(\pi\) to achieve a targeted distribution of stances, \(D\), and operates strictly as a content-agnostic mechanism. This means \(f\) exclusively uses relational properties both as input features and during the learning process, without access to any intrinsic properties.
\end{definition}

\begin{theorem}[Non-Determinacy of Content-Agnostic Moderation]\label{thm:non_determinacy}
    It is not guaranteed that a content-agnostic moderation function \(f\) can achieve a targeted distribution \(D\) with only relational properties available.
\end{theorem}

This is supported by two proofs outlined below, details in App.~\ref{app:proofs}.
\vspace{-5mm}
\begin{proof}[Proof by contradiction]
    Suppose the only available features for a moderation function \(f\) are relational, then we can always construct two recommendations \(\pi'\) and \(\pi''\) that are indistinguishable as input to \(f\), yet require different moderations to achieve \(D\). 
\end{proof}
\vspace{-5mm}
\begin{proof}[Proof from learning uncertainty]
    Training a moderation function \(f\) to consistently produce an output \(\pi'\) aligning with \(D\) is unfeasible if the only available features for the learning algorithm are relational, because there exist an output \(\pi''\) indistinguishable in relational properties from \(\pi'\) yet with \(D' \neq D\).
\end{proof}

While Theorem \ref{thm:non_determinacy} indicates that content-agnostic moderation cannot be fundamentally guaranteed, this does not preclude its practical feasibility. In real-world settings, specific methods may achieve a target distribution under plausible assumptions as explored below.

\begin{table}[!h]
    \footnotesize
    \captionsetup{font=small}
    \centering
    \caption{Chosen popularity suppression methods.}
    \label{tab:popularity_suppression_methods}
    \begin{tabular}{llll}
    \toprule
       & Metric        & Agg-dimension & Thresholding \\
    \midrule
    KC & Probabilistic  & Global      & Soft         \\
    RR & Deterministic & Temporal    & Hard  \\
    \bottomrule      
    \end{tabular}
\end{table}

\subsection{Empirical Possibility of Content-Agnostic Moderation}\label{sec:empirical_possibility}
We consider two potential working conditions of content-agnostic moderation.
First, if the moderation function can infer item stances from relational properties, it may achieve the desired stance distribution using this proxy information.
Second, if the target distribution is uniform across stances and the data is favorably balanced, a content-agnostic moderation function could attain neutrality.

\subsubsection{Proxy-Based Moderation}\label{sec:theory_proxy_based}
This method leverages inferred correlations between user preferences and item content to develop proxies for stances from the recommendation configuration. By transporting the proxy distribution, the moderation function tries to achieve stance neutrality.

A key assumption is that user behaviors, which reflect stance leaning, can indirectly indicate the stances of the items themselves. This assumption is grounded in the understanding that recommendation systems aim to learn from and predict user behaviors, which include preferences related to item stances. This implies that the moderation's effectiveness hinges on the upstream recommender's ability to accurately capture user preferences and item stances.

\subsubsection{Egalitarian Exposure Enforcement}\label{sec:theory_uniform_exposure}

Now we examine a more specific problem setting where the desired distribution is \emph{uniform} across stances. This means stance-neutral exposure in the recommendation setting, i.e. items from each stance in a recommendation configuration have equal exposure, both globally across all users and locally within each user group, without considering the rank or order of items within the recommendation lists.

Formally, the \emph{exposure} \(e_i\) of an item \(i \in \mathcal{I}\) is \(e_i = \sum_{u \in \mathcal{U}} \text{count}(i, \pi_u)\), with \(\text{count}(i, \pi_u)\) representing how often item \(i\) appears in the recommendation list \(\pi_u\). The objective is uniform stance distribution across items, formulated as:
\[
\forall s \in S, \quad \frac{\sum_{i \in \mathcal{I}_s} e_i}{\sum_{j \in \mathcal{I}} e_j} = \frac{1}{|S|}
\]
where \(\mathcal{I}_s\) denotes the items in \(\mathcal{I}\) with stance \(s\).

Under this problem setting, the Egalitarian Exposure Enforcement strategy could work with favorable data distribution. This strategy ensures that every item \(i \in \mathcal{I}\) receives equal exposure across all users in \(\mathcal{U}\), formally defined by the condition \(e_i = e_j\) for all \(i, j \in \mathcal{I}\). It can be shown (see Appx.~\ref{app:proofs}) that if the pool of items is uniformly distributed across stances, Egalitarian Exposure Enforcement will lead to stance neutrality over time.

\section{Experiment settings}\label{sec:experiment}
We first introduce our simulation framework in Sec.~\ref{sec:simulation}, followed by descriptions of each component: recommenders in Sec.~\ref{sec:rec_models}, moderators in  Sec.~\ref{sec:moderators} and users in Sec.~\ref{sec:user_models}, as well as the semi-synthetic data generation (Sec.~\ref{sec:data}). We then define the evaluation metrics in Sec.~\ref{sec:evaluation_metrics}. The simulation environment will then be used in Sec.~\ref{sec:results} to address the research questions.

\subsection{Notations}\label{sec:notations}
Let \( m \) and \( n \) be the number of users and items, respectively. \( S \) is the set of all possible stances, with \( |S| \) being the number of distinct stances. \( T \) and \( k \) respectively denote the total number of time steps and recommendations per user at each step.
\( R \) is a general term for the discrete form of recommendation, with \( R_{ij} \) being the item \( j \) recommended to user \( i \).
\({RS}_t \) aggregates sequences of recommendations for each user at time \( t \), and \({RM}_t \in \{0, 1\}^{m \times n} \) is the corresponding binary matrix, which ignores the order of recommendations.
\({RS}_{\text{agg},t} \) and \({RM}_{\text{agg},t} \in \mathbb{N}^{m \times n} \) compile these sequences and matrices up to time \( t \).
The above recommendation configurations are relational, to be used by content-agnostic moderation functions.
Matrices \( U \) and \( A \), sized \( m \times |S| \) and \( n \times |S| \), represent user preferences and item stances, respectively. User groups are segmented by stance, labeled \( UC_s \) for each stance \( s \in S\). Clearly, \(U\) and \(A\) are intrinsic properties, forbidden to be used by any content-agnostic moderation method.

\subsection{Simulated Feedback Loop}\label{sec:simulation}
We set up a simulation feedback loop iterates through a recommendation model and the users, potentially mediated by a content-agnostic moderator as shown in Fig.~\ref{fig:feedback_loop}. At each time step $t$, the \textbf{recommender} outputs an ordered list of \( k \) items for each user. With moderation applied, the \textbf{moderator} adjusts this list and outputs a revised set of top-\( k \) items. \textbf{Users} then receive these items and decide whether to consume them based on their preferences, which further get updated. The users' behavior data informs the next iteration of the recommendation model. 
Interaction histories are accumulated over \( T \) time steps. For our experiments, we set \( k = 5 \) and \( T = 60 \), due to the limited data size.\footnote{Our simulation framework bears some similarity with \cite{liu2021interaction} and \cite{zhang2023evolution} with important modifications listed in Appendix~\ref{app:experiment_settings}. }

\subsubsection{Recommendation Models}\label{sec:rec_models}
To address \ref{rq1}, we employ two classical content-unaware recommendation models: Matrix Factorization (MF) and Popularity-based (PP), relying solely on relational properties, specifically user-item interactions, as input.\footnote{While incorporating other intrinsic properties such as user demographics or news sources is plausible, we reserve such extensions for future work.}
These models are compared with Oracle-CB, an idealized content-based model. Oracle-CB simplifies the complexity typically associated with content-based models by using the user preference matrix \(U\) and item stance matrix \(A\) to generate the top-\(k\) items for each user through a straightforward dot product of \(U\) and \(A\). This abstraction from specific model details allows us to avoid the intricate architectures and content processing typical of many content-based models, which would introduce unnecessary complications into our study. Despite its unrealistic assumption of knowing the latent user preferences and item stances, Oracle-CB serves as a useful benchmark to evaluate the theoretical ideal of content-based recommendation.

To address \ref{rq2}, we examine content-based (CB) models with varying levels of accuracy by introducing randomness into the models. We categorize these levels as perfect (Oracle-CB), moderate (Inaccurate-CB), and low (Random Recommendation). This stratification allows us to focus primarily on the effects of moderation, minimizing distractions from the complexities of model architecture and training. Specifically, we aim to understand how the assumptions of moderation strategies influence these effects.

As noted in \ref{sec:theory_proxy_based}, the success of proxy-based moderation strategies depends on the recommender's accuracy in capturing user preferences and item stances; therefore, we expect better performance from Oracle-CB due to its high accuracy, with likely declines in performance for Inaccurate-CB and Random Recommendation.

To address \ref{rq3} concerning users' change of opinions, we fix the recommender to be Oracle-CB to isolate any effects that might stem from differing model behaviors.

\subsubsection{Moderators}\label{sec:moderators}
As discussed in Sec.~\ref{sec:empirical_possibility}, we examine two content-agnostic moderation strategies that use only the ordered lists \(RS_t\) from the recommender, without extra user or item details. The main objective is to even out the stance distribution of items shown to users, potentially reducing opinion polarization.

\emph{Egalitarian Exposure Enforcement.} Used in both academic and industrial settings, this approach aims to either suppress popularity or enhance diversity. The common idea involves thresholding popularity metrics either probabilistic (score-based) or deterministic (count-based), applied globally or temporally. Thresholding ranges from soft (lowering the likelihood of recommending popular items) to hard (excluding items that exceed a certain quota).

We examine two methods: Knapsack Constrained Optimization (KC) and Round-Robin procedure (RR), categorized in Table~\ref{tab:popularity_suppression_methods}. \emph{KC} uses mixture integer programming for soft thresholding to cap global item popularity at each time step, defined as:
\begin{equation}\label{eq:kc_constraint}
    \lambda \cdot \sum_{i=1}^{m} \sum_{j=1}^{n} RM_{t, ij} \left(\frac{\sum_{i=1}^{m} RM_{t, ij}}{\sum_{i=1}^{m} \sum_{j=1}^{n} RM_{t, ij}}\right)
\end{equation}
Here, \(\lambda \in (0, 1]\) is a hyperparameter that controls moderation strength---the lower \(\lambda\), the stronger the moderation. The second term represents the total popularity of items from the original recommendations at time step \(t\).
\emph{RR} applies hard thresholding by assigning equal exposure quotas to each item over a defined period. It selects the top-preferred item for each user in each round unless the item’s quota is exceeded. We developed a self-adjusting quota mechanism that dynamically sets quotas as the minimum integer value needed to ensure each user receives \(k\) items without violating exposure equality.
While other combinations are feasible, they are omitted to narrow the focus of this study.

Note on implementations: Inspired by \cite{seymenunified2021} and \cite{ribeiro2012paretoefficient}, our KC algorithm uses a more flexible and readily adjustable hyperparameter, processes inputs based on recommendation counts, and optimizes the Hamming distance from the initial recommendations. Influenced by \cite{patro2020fairrec} and \cite{wu2021tfrom}, our RR approach introduces a self-adjusting quota mechanism is tailored to moderating recommendation counts. Detailed descriptions are provided in Appendix~\ref{app:moderators}.

\begin{algorithm}[ht]
    \SetAlgoHangIndent{0.5em}
    \begin{small}
    \captionsetup{font=small}
    \caption{Cluster Dispersal Algorithm}
    \label{alg:cluster_dispersal}

    \SetKwInput{KwInput}{Input}                %
    \SetKwInput{KwOutput}{Output}              %

    \KwInput{$RS_t$: recommendation lists at $t$, $RM_{\text{agg},t}$: aggregated recommendation matrix up to $t$, number of items to be swapped per user $\alpha$, Silhouette score threshold $\beta$}
    \KwOutput{Modified recommendation $RS_t$}

    $C \leftarrow \text{Coclustering}(RM_{\text{agg},t})$  
    \tcp{Find user-item co-clusters} 
    $TightClusters \leftarrow \text{IdentifyTightClusters}(C, \beta)$\;  %
    
    \For{$u \in TightClusters$}{  %
        $BottomItems \leftarrow \text{GetBottomItems}(RS_{t,u}, \alpha)$\;  %
        \uIf{Method is RD}{
            $SwapItems \leftarrow \text{RandomSample}(\text{OtherClustersItems}(C, u), \alpha)$\;  
        }
        \uElseIf{Method is SD}{
            $SwapItems \leftarrow \text{SimilarityBasedSample}(\text{OtherClustersItems}(C, u), \alpha)$\;  
        }
        $RS_{t,u} \leftarrow (RS_{t,u} \setminus BottomItems) \cup SwapItems$\;  %
    }
    
    \Return{$RS_t$}\;  %
    \end{small}
\end{algorithm}

\emph{Proxy-Based Moderation}
We propose \textbf{Cluster Dispersal}, a novel content-agnostic moderation strategy using biclustering of users and items based only on the aggregated recommendation matrix. The aim is to adjust recommendation lists to reduce the frequency of overly popular stances and increase exposure to minority stances. This is achieved through two methods: Random Dispersal (RD) and Similarity-Based Dispersal (SD), as outlined in Algorithm~\ref{alg:cluster_dispersal}.

Both RD and SD start by performing spectral decomposition on the aggregated recommendation matrix to bicluster users and items, identifying clusters where the overall Silhouette score surpasses a predefined threshold \(\beta\). This threshold determines the tightness of clusters warranting dispersal. For users in dense clusters, RD swaps each user’s \(\alpha\) least preferred items from the original recommendation with random alternatives from other clusters. In contrast, SD selects replacement from other clusters whose spectral embeddings, derived from decomposition, close to the user's.

\emph{Hyperparameters for Moderation Strength}\label{sec:hyperparameters}
For KC, a smaller $\lambda$ implies stronger moderation, with values tested at $\lambda=0.6$ and $0.4$. For RD and SD, $\alpha$ dictates the number of items swapped per user, tested at $\alpha=1$ and $2$. The threshold $\beta$, set at $0.45$ based on initial explorations, indicates the point at which cluster dispersal is triggered. This subtler control of moderation is critical for fine-tuning and is left for further investigation. 
To facilitate direct comparisons across moderation methods, we first choose $\alpha=1$ for our cluster dispersal methods for the least drops in CTR, and then choose the $\lambda$ for KC to make the drop in CTRs closest to the cluster dispersal methods (Table~\ref{tab:rq2_9t}, \ref{tab:rq2_9f}, \ref{tab:rq2_2} and \ref{tab:rq2_0}). We also plot the Pareto frontiers from varied levels of moderation strengths (Fig.~\ref{fig:pareto}).

\subsubsection{User Models}\label{sec:user_models}
User models are needed for completing the simulation loop. In our simulations, users select a subset of items to read from those shown, based on a similarity-based user \textbf{choice} model adding slight random noise. Users' preferences are subject to influence by what they read, although the extent and nature of this influence can vary. We use a basic linear model\footnote{Other more complicated opinion updating models are possible, especially in computational social science, such as \cite{bolzern2019opinion} and \cite{perra2019modelling}.} to \textbf{update} user preferences each time a user \(u\) reads (clicks on) an item \(i\), as described by the formula \cite{zhang2023evolution}:
\begin{equation}
    U_u \leftarrow U_u + \gamma A_i \label{eq:user_update}
\end{equation}
Here, \(A_i\) represents the stance vector of item \(i\), and \(\gamma\) is a small constant that quantifies the influence of read content on user preferences. For our experiments addressing \ref{rq1} and \ref{rq2}, we compare a non-influential (\(\gamma=0\)) against a moderate influence setting (\(\gamma=0.001\)). We explore the effects of larger values of \(\gamma\) for \ref{rq3}.

\subsection{Data}\label{sec:data}
We define stances \(S\) as Left (L), Center (C), and Right (R) for simplicity. \emph{User preferences}, represented by a real-valued matrix \(U\) of dimensions \(m \times 3\), are generated to reflect the U.S. political landscape based on the latest PEW survey \cite{doherty2021beyond}, with proportions \(UC_L/UC_C/UC_R = 56/11/33\). \emph{Items} are represented by a binary stance matrix \(A\) with dimensions \(n \times 3\), indicating the stance to which each item belongs.

We have developed multiple simulation scenarios to mirror real-world variations in content distribution and user behavior, assessing the data sensitivity of different moderation strategies, especially the Egalitarian Exposure Enforcement strategy.

\emph{Scenario 1: Balanced Baseline}
Items are equally distributed among the three stances, with \(\frac{n}{3}\) items for each, serving as a baseline to assess moderation techniques in a neutral environment.

\emph{Scenario 2: Bi-Polarized Content}
A polarized distribution with \(0.45n\) items for L and R, and just \(0.1n\) for C, mirrors a divided media landscape where centrist views are underrepresented.

\emph{Scenario 3: Propaganda Dominance from the Left}
The content distribution is heavily biased towards Left with \(0.8n\) items, while Center and Right have only \(0.1n\) items each, akin to propagandistic or biased media conditions. The choice of Left as the dominant stance was made randomly and does not reflect any particular preference or significance.

\emph{Scenario 4: User Preference Amplification for Right Stance}
Right-leaning user preferences are artificially enhanced by scaling up their preference vectors, aiming to simulate the formation of opinion leaders.
Interestingly, we did not observe a notable difference in the performance of moderators from Scenario 1, suggesting that merely amplifying preferences might only indirectly affect user behavior. More direct interventions will be explored in future work.

As discussed in \ref{sec:theory_uniform_exposure}, Egalitarian Exposure Enforcement is most effective when stances are evenly distributed across the item pool, as seen in Scenarios 1 and 4. Scenario 3 presents the most challenging conditions. The impact on Scenario 2 is less certain; although it is bi-polarized, the equal representation of Left and Right might still balance the overall stance distribution.

In each scenario, we set \(m=100\) and \(n=3000\). The simulation begins with a \emph{bootstrapping} phase where each user is presented with 10 random articles, and interactions are recorded based on the user choice model described in Sec.~\ref{sec:user_models}. To address the issue of cold items, any item without initial interaction is shown to 10 random users until fewer than 0.5\% of items remain unselected. This ensures widespread initial exposure and establishes an effective interaction matrix at \(t_0\), ready for matrix factorization and other decomposition techniques.

\begin{table*}[ht]
    \centering
    \setlength{\tabcolsep}{10pt}
    \small
    \resizebox{0.9\textwidth}{!}{
    \begin{threeparttable}
    \caption{Comparison of recommendation models without moderation on CTR and JSD.}
    \begin{tabular}{llllllll}
        \toprule
        Scenario & User update & Model & CTR & JSD-O & JSD-G & JSD-O (read) & JSD-G (read) \\
        \midrule
        \multirow[t]{6}{*}{S1. Balanced baseline} & \multirow[t]{3}{*}{False} & Oracle-CB & 0.812 & 0.189 & 0.456 & 0.212** & 0.506** \\
         &  & MF & 0.573 (-29.4\%**) & 0.092** & 0.131** & 0.135** & 0.238** \\
         &  & PP & \textit{0.392 (-51.7\%**)} & 0.182ns & 0.282** & \underline{0.247**} & 0.276** \\
        \cline{2-8}
         & \multirow[t]{3}{*}{True} & Oracle-CB & 0.879 & 0.201 & 0.481 & 0.219** & 0.521** \\
         &  & MF & 0.602 (-31.5\%**) & 0.092** & 0.129** & 0.135** & 0.233** \\
         &  & PP & \textit{0.413 (-53.0\%**)} & 0.18ns & 0.275** & \underline{0.249**} & 0.276** \\
        \cline{1-8} 
        \multirow[t]{6}{*}{S2. Bi-polarized content} & \multirow[t]{3}{*}{False} & Oracle-CB & 0.813 & 0.257 & 0.408 & 0.263** & 0.457** \\
         &  & MF & \textit{0.434 (-46.6\%**)} & 0.23** & 0.252** & 0.232ns & 0.271** \\
         &  & PP & \textit{0.363 (-55.3\%**)} & 0.246ns & 0.338** & \underline{0.29**} & 0.328** \\
        \cline{2-8}
         & \multirow[t]{3}{*}{True} & Oracle-CB & 0.876 & 0.257 & 0.422 & 0.264** & 0.468** \\
         &  & MF & \textit{0.459 (-47.6\%**)} & 0.227** & 0.249** & 0.232ns & 0.269** \\
         &  & PP & \textit{0.39 (-55.5\%**)} & 0.246ns & 0.333** & \underline{0.295**} & 0.33** \\
        \cline{1-8} 
        \multirow[t]{6}{*}{S3. Propaganda dominance} & \multirow[t]{3}{*}{False} & Oracle-CB & 0.802 & 0.249 & 0.421 & 0.257** & 0.467** \\
         &  & MF & \textit{0.48 (-40.1\%**)} & \underline{0.344**} & 0.395** & 0.401** & 0.399** \\
         &  & PP & 0.522 (-34.9\%**) & \underline{0.421**} & 0.46** & \underline{0.451**} & \underline{0.469**} \\
        \cline{2-8}
         & \multirow[t]{3}{*}{True} & Oracle-CB & 0.866 & 0.255 & 0.439 & 0.263** & 0.477** \\
         &  & MF & \textit{0.509 (-41.2\%**)} & \underline{0.335**} & 0.386** & 0.399** & 0.398** \\
         &  & PP & 0.557 (-35.6\%**) & \underline{0.413**} & \underline{0.453*} & \underline{0.45**} & 0.466** \\
        \cline{1-8}
        \multirow[t]{6}{*}{S4. Amplified right-stance} & \multirow[t]{3}{*}{False} & Oracle-CB & 0.843 & 0.195 & 0.465 & 0.214** & 0.51** \\
         &  & MF & \textit{0.495 (-41.2\%**)} & \textbf{0.109**} & \textbf{0.121**} & 0.152** & \textbf{0.177*} \\
         &  & PP & \textit{0.485 (-42.4\%**)} & \underline{0.354**} & 0.393** & \underline{0.37**} & 0.383** \\
        \cline{2-8}
         & \multirow[t]{3}{*}{True} & Oracle-CB & 0.891 & 0.202 & 0.485 & 0.219** & 0.524** \\
         &  & MF & \textit{0.52 (-41.7\%**)} & \textbf{0.104**} & \textbf{0.116**} & 0.149** & \textbf{0.173**} \\
         &  & PP & \textit{0.515 (-42.2\%**)} & \underline{0.347**} & 0.385** & \underline{0.367**} & 0.38** \\
        \bottomrule
        \end{tabular}  
    \begin{tablenotes}[flushleft, para, wide]
        \small
        \item[] Note: The first number in each cell represents the absolute value; the percentage in parentheses indicates the change from the Oracle-CB for CTR. The interaction matrix at $t_0$ is the results of bootstrapping using random exposure as described in Sec.~\ref{sec:data}, hence only the JSD-O (read) and JSD-G (read) are available for comparison with $t_0$, while the t-tests for JSD-O and JSD-G of the shown items are performed between Oracle-CB and the other two content-unaware models MF and PP. The worsened JSDs, comparing with the content-based model, are \underline{underlined}. Great decreasings in JSDs are \textbf{boldfaced}. CTR drops over 40\% are \textit{italicized}.
        \item[] *: \( p \)-value < 0.05, significant; **: \( p \)-value < 0.01, very significant; ns: \( p \)-value > 0.05, not significant.
    \end{tablenotes}
    
    \label{tab:rq1}
    \end{threeparttable}
    }
    \end{table*}
\subsection{Evaluation Metrics}\label{sec:evaluation_metrics}
We use the click-through rate (CTR), a standard measure of recommendation accuracy. For assessing stance neutrality, we apply two Jensen-Shannon distance-based metrics. Additionally, we use two metrics based on user preferences to quantify opinion dynamics:

\textbf{JSD-O (Jensen-Shannon Distance - Overall)} quantifies the disparity between the distribution of stances in shown or read items, \(P\), and a uniform stance distribution, \(Q\). It is defined as the square root of the Jensen-Shannon divergence between \(P\) and \(Q\) \cite{lin1991divergence}:
\[
\text{JSD-O}(P \| Q) = \sqrt{\frac{1}{2} KL(P \| M) + \frac{1}{2} KL(Q \| M)}
\]
\(M = \frac{1}{2}(P + Q)\) and \(KL\) denotes the Kullback-Leibler divergence.

\textbf{JSD-G (Jensen-Shannon Distance - User Group Specific)} calculates the average Jensen-Shannon distance across user groups. It measures the distance between the stance distribution \(PU_s\) of items for each group $UC_s$ and the uniform distribution \(Q\):
\[
\text{JSD-G} = \frac{1}{{|S|}} \sum_{s \in S} \text{JSD}(PU_s \| Q)
\]
Each individual JSD is computed in the same manner as JSD-O.

A \textbf{stance-neutral} recommendation system achieves a JSD-O of zero, indicating complete exposure neutrality, and a JSD-G of zero, signifying unbiased group-specific item exposure.

We primarily focus on the JSD metrics for items shown by the recommender to evaluate its neutrality. We also examine the JSD metrics for items users actually read, referred to as ``JSD-* (read)'', to gain insights into users' opinion dynamics.

\textbf{UMOE (User Mean Opinion Extremeness)} captures the polarization of users' stances by calculating the variance of each user's preference strength across the three stances. A higher variance indicates more extreme stances. UMOE averages the variances across all users.
\textbf{UMS (User Mean Stance)}:
To translate users' preferences across three stances into a scalar value, we assign numerical values \(S = [-1, 0, 1]\) to stances L/C/R, respectively. Each user's stance, \(S_u\), is calculated as:
\(
S_u = \sum_{i=1}^3 U_u[i] \cdot S[i]
\),
where \(U_u\) is the preference vector for user \(u\). UMS averages \(S_u\) across all users to provide a measure of the overall leaning of the population.

\begin{table*}[hbt]
    \setlength{\tabcolsep}{3pt}
    \small
    \resizebox{0.9\textwidth}{!}{
    \begin{threeparttable}
    \caption{Comparison of moderation methods with \emph{Oracle-CB} recommender with user preference \emph{updating}.}
    \label{tab:rq2_9t}
    \begin{tabular}{lllllllllllll}
        \toprule
         & \multicolumn{3}{r}{S1. Balanced baseline} & \multicolumn{3}{r}{S2. Bi-polarized content} & \multicolumn{3}{r}{S3. Propaganda dominance} & \multicolumn{3}{r}{S4. Amplified right-stance} \\
         & CTR & JSD-O & JSD-G & CTR & JSD-O & JSD-G & CTR & JSD-O & JSD-G & CTR & JSD-O & JSD-G \\
        \midrule
        - & 0.879 & 0.201 & 0.481 & 0.876 & 0.257 & 0.422 & 0.866 & 0.255 & 0.439 & 0.891 & 0.202 & 0.485 \\
        RR & \textit{0.661 (-24.8\%**)} & 0.109** & \textbf{0.294**} & \textit{0.672 (-23.3\%**)} & 0.214** & 0.275** & \textit{0.671 (-22.5\%**)} & \underline{0.364**} & 0.385** & \textit{0.673 (-24.5\%**)} & \textbf{0.108**} & \textbf{0.295**} \\
        KC & 0.762 (-13.3\%**) & \textbf{0.108**} & 0.338** & 0.759 (-13.3\%**) & 0.19** & \textbf{0.263**} & 0.746 (-13.9\%**) & \underline{0.298**} & 0.344** & 0.773 (-13.3\%**) & \textbf{0.108**} & 0.338** \\
        RD & 0.767 (-12.7\%**) & 0.131** & 0.342** & 0.752 (-14.2\%**) & 0.228** & 0.294** & 0.767 (-11.4\%**) & 0.22** & \textbf{0.328**} & 0.78 (-12.5\%**) & 0.136** & 0.348** \\
        SD & \textbf{0.772 (-12.1\%**)} & 0.126** & 0.357** & \textbf{0.773 (-11.7\%**)} & \textbf{0.184**} & 0.306** & \textbf{0.776 (-10.4\%**)} & \textbf{0.188**} & \textbf{0.328**} & \textbf{0.784 (-12.1\%**)} & 0.136** & 0.36** \\
        \bottomrule
        \end{tabular}
    
    \begin{tablenotes}[flushleft, para, wide]
        \small
        \item Note: Best results are \textbf{boldfaced} with either least drops in CTR or smallest JSD values. The worsened JSDs, comparing with no moderation, are \underline{underlined}. CTR drops over 20\% are \textit{italicized}.
        \end{tablenotes}
    \end{threeparttable}
    }
    \end{table*}

\begin{table*}[hbt]
    \setlength{\tabcolsep}{3pt}    
    \begin{small}
    \caption{Comparison of moderation methods with \emph{Oracle-CB} recommender \emph{without} user preference updating.}
    \resizebox{0.9\textwidth}{!}{
    \begin{tabular}{lllllllllllll}
        \toprule
         & \multicolumn{3}{r}{S1. Balanced baseline} & \multicolumn{3}{r}{S2. Bi-polarized content} & \multicolumn{3}{r}{S3. Propaganda dominance} & \multicolumn{3}{r}{S4. Amplified right-stance} \\
         & CTR & JSD-O & JSD-G & CTR & JSD-O & JSD-G & CTR & JSD-O & JSD-G & CTR & JSD-O & JSD-G \\
        \midrule
        - & 0.812 & 0.189 & 0.456 & 0.813 & 0.257 & 0.408 & 0.802 & 0.249 & 0.421 & 0.843 & 0.195 & 0.465 \\
        RR & \textit{0.627 (-22.8\%**)} & 0.115** & \textbf{0.29**} & \textit{0.636 (-21.8\%**)} & 0.219** & 0.277** & \textit{0.632 (-21.2\%**)} & \underline{0.368**} & 0.388** & \textit{0.651 (-22.7\%**)} & 0.108** & \textbf{0.292**} \\
        KC & 0.713 (-12.1\%**) & \textbf{0.106**} & 0.324** & 0.713 (-12.4\%**) & 0.189** & \textbf{0.255**} & 0.698 (-12.9\%**) & \underline{0.297**} & 0.336** & 0.734 (-12.9\%**) & \textbf{0.106**} & 0.324** \\
        RD & 0.719 (-11.4\%**) & 0.125** & 0.331** & 0.707 (-13.0\%**) & 0.227** & 0.289** & 0.719 (-10.3\%**) & 0.216** & 0.319** & 0.744 (-11.7\%**) & 0.132** & 0.339** \\
        SD & \textbf{0.724 (-10.8\%**)} & 0.122** & 0.342** & \textbf{0.726 (-10.7\%**)} & \textbf{0.185**} & 0.297** & \textbf{0.725 (-9.5\%**)} & \textbf{0.186**} & \textbf{0.316**} & \textbf{0.748 (-11.2\%**)} & 0.132** & 0.347** \\
        \bottomrule
        \end{tabular}
    }
    \label{tab:rq2_9f}
    \end{small}
    \end{table*}
\begin{table*}[hbt]
    \setlength{\tabcolsep}{3pt}
    \begin{small}
    \caption{Comparison of moderation methods with \emph{inaccurate-CB} recommender with user preference \emph{updating}.}
    \resizebox{0.9\textwidth}{!}{
    \begin{tabular}{lllllllllllll}
        \toprule
         & \multicolumn{3}{r}{S1. Balanced baseline} & \multicolumn{3}{r}{S2. Bi-polarized content} & \multicolumn{3}{r}{S3. Propaganda dominance} & \multicolumn{3}{r}{S4. Amplified right-stance} \\
         & CTR & JSD-O & JSD-G & CTR & JSD-O & JSD-G & CTR & JSD-O & JSD-G & CTR & JSD-O & JSD-G \\
        \midrule
        - & 0.667 & 0.05 & 0.167 & 0.708 & 0.226 & 0.259 & 0.671 & 0.337 & 0.327 & 0.681 & 0.049 & 0.172 \\
        RR & 0.58 (-13.1\%**) & \underline{0.069*} & 0.15* & 0.595 (-16.0\%**) & 0.204** & 0.228** & 0.628 (-6.4\%**) & \textbf{0.326*} & \textbf{0.32ns} & 0.591 (-13.2\%**) & \underline{0.064*} & 0.149** \\
        KC & \textbf{0.616 (-7.6\%**)} & \underline{0.091**} & 0.167ns & 0.641 (-9.4\%**) & \textbf{0.159**} & \textbf{0.191**} & 0.63 (-6.1\%**) & \underline{0.365**} & \underline{0.347**} & 0.623 (-8.4\%**) & \underline{0.09**} & 0.17ns \\
        RD & 0.613 (-8.2\%**) & \textbf{0.028**} & \textbf{0.123**} & \textbf{0.645 (-8.8\%**)} & 0.22ns & 0.238** & 0.652 (-2.8\%**) & 0.344ns & 0.331ns & \textbf{0.626 (-8.1\%**)} & \textbf{0.034**} & \textbf{0.128**} \\
        SD & 0.612 (-8.2\%**) & 0.04ns & 0.132** & 0.641 (-9.4\%**) & 0.233ns & 0.252ns & \textbf{0.652 (-2.8\%**)} & 0.344ns & 0.333ns & 0.621 (-8.8\%**) & 0.055ns & 0.137** \\
        \bottomrule
    \end{tabular}
    }
    \label{tab:rq2_2}
    \end{small}
    \end{table*}
\begin{table*}[hbt]
    \setlength{\tabcolsep}{3pt}
    \begin{small}    
    \caption{Comparison of moderation methods with \emph{random recommender} with user preference \emph{updating}.}
    \resizebox{0.9\textwidth}{!}{
    \begin{tabular}{lllllllllllll}
        \toprule
         & \multicolumn{3}{r}{S1. Balanced baseline} & \multicolumn{3}{r}{S2. Bi-polarized content} & \multicolumn{3}{r}{S3. Propaganda dominance} & \multicolumn{3}{r}{S4. Amplified right-stance} \\
         & CTR & JSD-O & JSD-G & CTR & JSD-O & JSD-G & CTR & JSD-O & JSD-G & CTR & JSD-O & JSD-G \\
         \midrule
        - & 0.522 & 0.02 & 0.041 & 0.516 & 0.205 & 0.209 & 0.591 & 0.34 & 0.344 & 0.527 & 0.02 & 0.041 \\
        RR & 0.501 (-3.9\%ns) & 0.062** & 0.096** & 0.498 (-3.6\%ns) & 0.214ns & 0.222ns & 0.567 (-4.1\%*) & \textbf{0.309**} & \textbf{0.315**} & 0.507 (-3.8\%ns) & 0.057** & 0.093** \\
        KC & \textbf{0.512 (-1.9\%ns)} & 0.08** & 0.108** & \textbf{0.504 (-2.2\%ns)} & \textbf{0.137**} & \textbf{0.16**} & 0.57 (-3.4\%*) & 0.342ns & 0.35ns & \textbf{0.517 (-2.0\%ns)} & 0.08** & 0.107** \\
        RD & 0.502 (-3.8\%*) & \textbf{0.021ns} & \textbf{0.042ns} & 0.501 (-3.0\%ns) & 0.206ns & 0.21ns & 0.574 (-2.8\%ns) & 0.34ns & 0.344ns & 0.507 (-3.7\%*) & \textbf{0.023ns} & \textbf{0.044ns} \\
        SD & 0.488 (-6.5\%**) & 0.045** & 0.059** & 0.489 (-5.3\%**) & 0.208ns & 0.213ns & \textbf{0.574 (-2.8\%ns)} & 0.341ns & 0.345ns & 0.494 (-6.2\%**) & 0.046** & 0.06** \\
        \bottomrule
    \end{tabular}
    }
    \label{tab:rq2_0}
    \end{small}
    \end{table*}
\section{Experimental Results}\label{sec:results}
This section presents the findings aligned with our research questions. Section~\ref{sec:rq1_results} explores the accuracy and stance neutrality of various recommenders without moderation, across diverse scenarios and user model updates, addressing \ref{rq1}. Section~\ref{sec:rq2_results} assesses the effectiveness of moderation strategies alongside recommenders of varying accuracy, tackling \ref{rq2}. Finally, Section~\ref{sec:rq3_results} investigates the impact of Cluster Dispersal moderation on user opinion dynamics with different moderation strengths, providing answers to \ref{rq3}.

\subsection{\ref{rq1}: Content-Unaware Recommenders}\label{sec:rq1_results}
\emph{CTR and JSD Analysis:}
Table~\ref{tab:rq1} shows that Oracle-CB consistently outperforms all other models in \emph{CTR} across scenarios, especially in settings with strong user preference signals such as S4. Content-unaware models like MF and PP show varying performances; MF generally outperforms PP except in heavily biased scenarios like S3. User updates tend to enhance CTRs across all models by amplifying preference signals.

\emph{JSD}-Os (Overall Stance Distribution) are typically lower than JSD-Gs (Group Specific), likely due to balancing effects from opposing user groups. JSDs for shown items are consistently lower than for read items, suggesting that consumption is more polarized than the recommendations themselves. Oracle-CB exhibits the highest JSD-O in scenarios S1 and S2, whereas PP performs worst in S3 and S4, significantly affected by propaganda and user preference amplification. MF's performance varies, likely due to its complex architecture and training dynamics.

\emph{Noteworthy Results:}
MF notably reduces JSD-O in S4, indicating its potential to balance stance distribution when minority group users are more active, as boldfaced in Table~\ref{tab:rq1}. However, reading behaviors remain polarized. Oracle-CB leads to considerable stance distortion across scenarios. Yet, in some cases, content-unaware models like PP and MF generate even higher JSDs than Oracle-CB (underlined in Table~\ref{tab:rq1}), showing that content-unaware models do \emph{not} inherently ensure stance neutrality.

\emph{Answering \ref{rq1}:}
Compared to Oracle-CB, content-unaware models like MF and PP sometimes achieve a more stance-neutral distribution but still contribute to significant polarization, particularly in biased media environments. Moreover, any neutralization achieved comes at the cost of markedly lower user engagement (CTRs often drop over 40\%, as italicized in Table \ref{tab:rq1}). This underscores the importance of employing moderation strategies that balance stance distribution without substantially compromising accuracy.
\begin{figure}
    \centering
    \captionsetup{font=footnotesize}
    \begin{subfigure}{0.495\linewidth}
        \includegraphics[width=\linewidth]{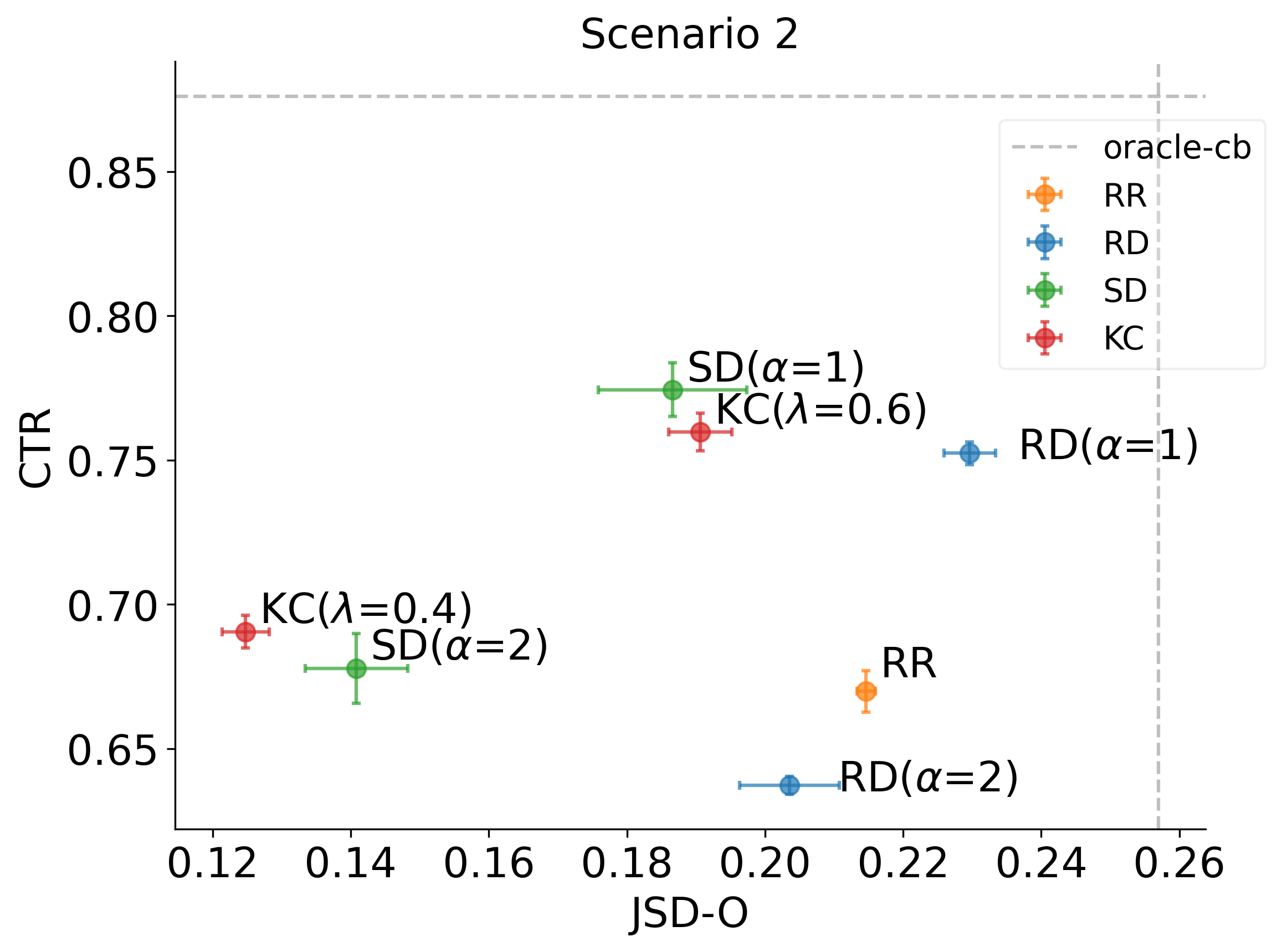}
    \end{subfigure}
    \hfill
    \begin{subfigure}{0.495\linewidth}
        \includegraphics[width=\linewidth]{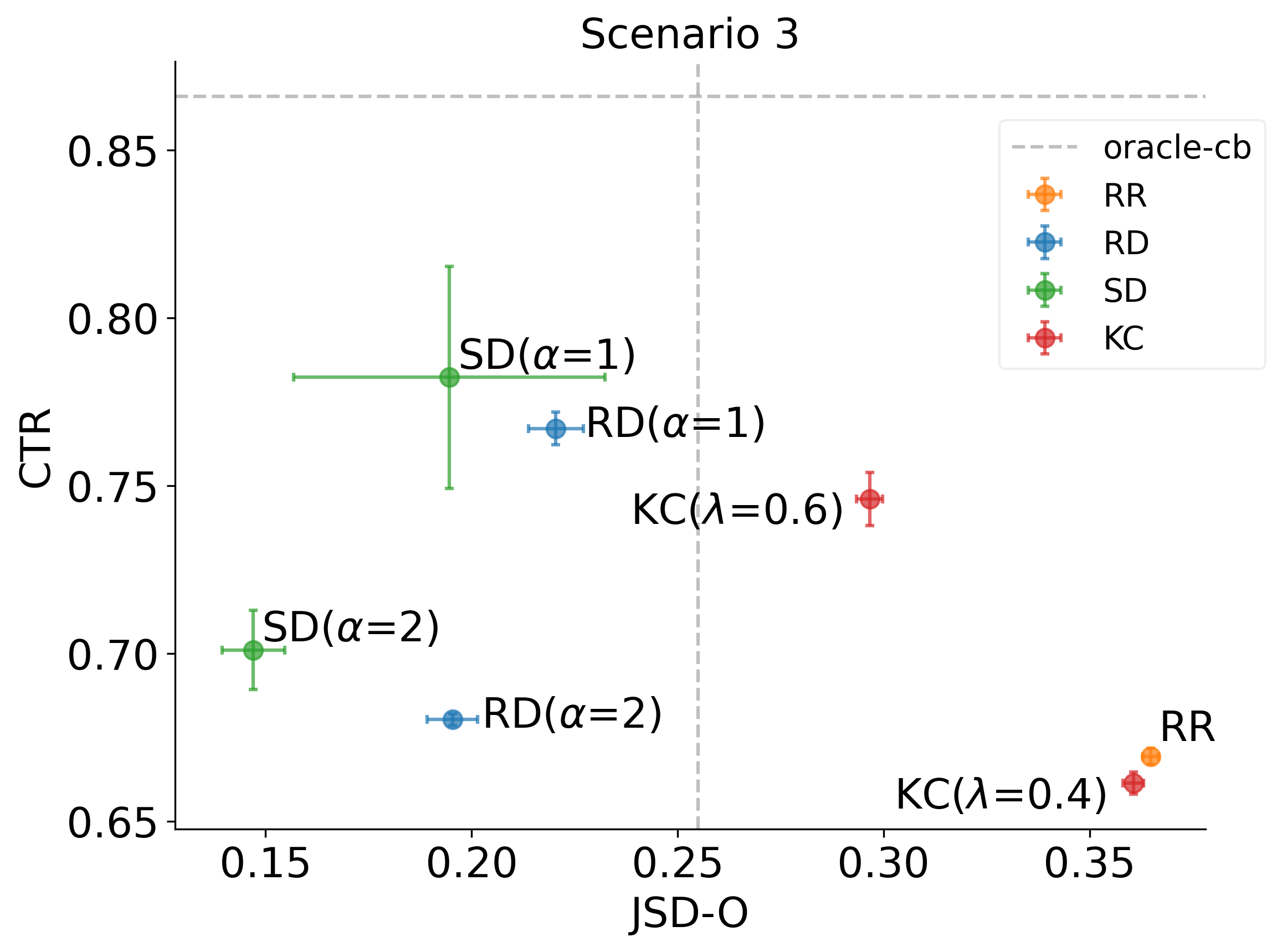}
    \end{subfigure}
    \caption{Pareto frontiers of different moderation strategies with \emph{Oracle-CB} recommender and user preference \emph{updating} in S2 and S3. Hyperparameters are bracketed, the meanings of which are explained in paragraph~\ref{sec:hyperparameters}. Markers are centered at the mean values over multiple runs, with errorbars roughly proportional to the standard deviations.}
    \label{fig:pareto}
\end{figure}
\subsection{\ref{rq2}: Content-Agnostic Moderation}\label{sec:rq2_results}
Tables \ref{tab:rq2_9t}, \ref{tab:rq2_9f}, \ref{tab:rq2_2}, and \ref{tab:rq2_0} present the results for evaluating various moderation strategies.
\begin{table*}[hbt]
    \setlength{\tabcolsep}{3pt}
    \begin{small}
    \caption{User effects comparison of moderation methods with \emph{Oracle-CB} recommender with user preference \emph{updating}.}
    \resizebox{1\textwidth}{!}{
    \begin{tabular}{lllllllllllllllll}
        \toprule
         & \multicolumn{4}{l}{S1. Balanced baseline} & \multicolumn{4}{l}{S2. Bi-polarized content} & \multicolumn{4}{l}{S3. Propaganda dominance} & \multicolumn{4}{l}{S4. Amplified right-stance} \\
         &                   UMS &     UMOE &       \makecell[l]{JSD-O\\(read)} &       \makecell[l]{JSD-G\\(read)} &                      UMS &     UMOE &       \makecell[l]{JSD-O\\(read)} &       \makecell[l]{JSD-G\\(read)} &                      UMS &     UMOE &       \makecell[l]{JSD-O\\(read)} &       \makecell[l]{JSD-G\\(read)} &                        UMS &     UMOE &       \makecell[l]{JSD-O\\(read)} &       \makecell[l]{JSD-G\\(read)} \\
        \midrule
        $t_0$ & -0.164 & 0.082 & 0.044 & 0.182 & -0.164 & 0.082 & 0.205 & 0.239 & -0.164 & 0.082 & 0.352 & 0.342 & -0.107 & 0.099 & 0.048 & 0.191 \\
        - & -0.190 & 0.115 & 0.219** & 0.521** & -0.191 & 0.114 & 0.264** & 0.468** & -0.199 & 0.113 & 0.263** & 0.477** & -0.135 & 0.134 & 0.219** & 0.524** \\
        RR & -0.155** & 0.093** & 0.108** & 0.386** & -0.158** & 0.099** & 0.213ns & 0.361** & -0.238** & 0.099** & 0.363ns & 0.375** & -0.101** & 0.110** & 0.108** & 0.388** \\
        KC & -0.194** & 0.103** & 0.15** & 0.433** & -0.200** & 0.104** & 0.222** & 0.383** & -0.225** & 0.105** & 0.292* & 0.409** & -0.138** & 0.120** & 0.147** & 0.436** \\
        RD & -0.181** & 0.105** & 0.164** & 0.453** & -0.181** & 0.106** & 0.24** & 0.406** & -0.195** & 0.106** & 0.237** & 0.424** & -0.125** & 0.124** & 0.168** & 0.458** \\
        SD & -0.180** & 0.106** & 0.16** & 0.463** & -0.180** & 0.106** & 0.209ns & 0.412** & -0.194** & 0.106** & 0.213** & 0.425** & -0.124** & 0.124** & 0.167** & 0.464** \\
        \bottomrule
        \end{tabular}
    }
    \label{tab:rq3_9t}
    \end{small}
\end{table*}
\emph{CTR and JSD Analysis:}
Moderation strategies introduce a trade-off between CTR and JSD. Among moderated systems, cluster-aware methods like RD and SD show smaller declines in CTR compared to baseline methods RR and KC. The effectiveness of content-agnostic moderation strategies varies with the accuracy of the underlying recommender. Less accurate models, such as random recommenders, exhibit smaller changes in JSD. In contrast, when paired with the more accurate Oracle-CB model, the benefits of moderation are more pronounced, especially with RD and SD.

\emph{Noteworthy Results:}
Cluster dispersal techniques consistently neutralize stances across different scenarios and outperform baseline methods, which sometimes cause significant CTR reductions (italicized in Tables \ref{tab:rq2_9t} and \ref{tab:rq2_9f}), and can exacerbate JSD-O, particularly in biased media environments (underlined in Tables \ref{tab:rq2_9t} and \ref{tab:rq2_9f}).

\emph{Pareto Frontiers Analysis:}
We visualize the trade-offs between CTR and JSD-O using Pareto frontiers for scenarios 2 and 3 (see Fig.~\ref{fig:pareto}), plots for other scenarios available in Appendix~\ref{app:pareto}. By adjusting the \(\alpha\) parameter, RD and SD allow for direct control over trade-offs, achieving desired outcomes effectively.
In contrast, KC's moderation strategy does not consistently yield the desired outcomes; a lower \(\lambda\) value leads to an increase in JSD-O in Scenario 3, showing that its adjustments can produce unpredictable results.

\emph{Answering \ref{rq2}:}
Our empirical findings demonstrate the superiority of the proxy-based cluster dispersal strategies RD and SD, which maintain accuracy while enhancing stance neutrality across various environments. These strategies surpass baseline methods RR and KC in terms of robustness in challenging conditions.  
Moreover, RD and SD offer significant computational advantages, operating at least 100 times faster than KC, enhancing the practical applicability of our novel methods (detailed in Appendix~\ref{app:additional_results}). 
The performance variances across data landscapes and recommender accuracy levels also underscore the impact of the assumptions made by moderation strategies, echoing our theoretical analysis in \ref{sec:empirical_possibility}.

\begin{figure}
    \centering
    \captionsetup{font=footnotesize}
    \includegraphics[height=48mm]{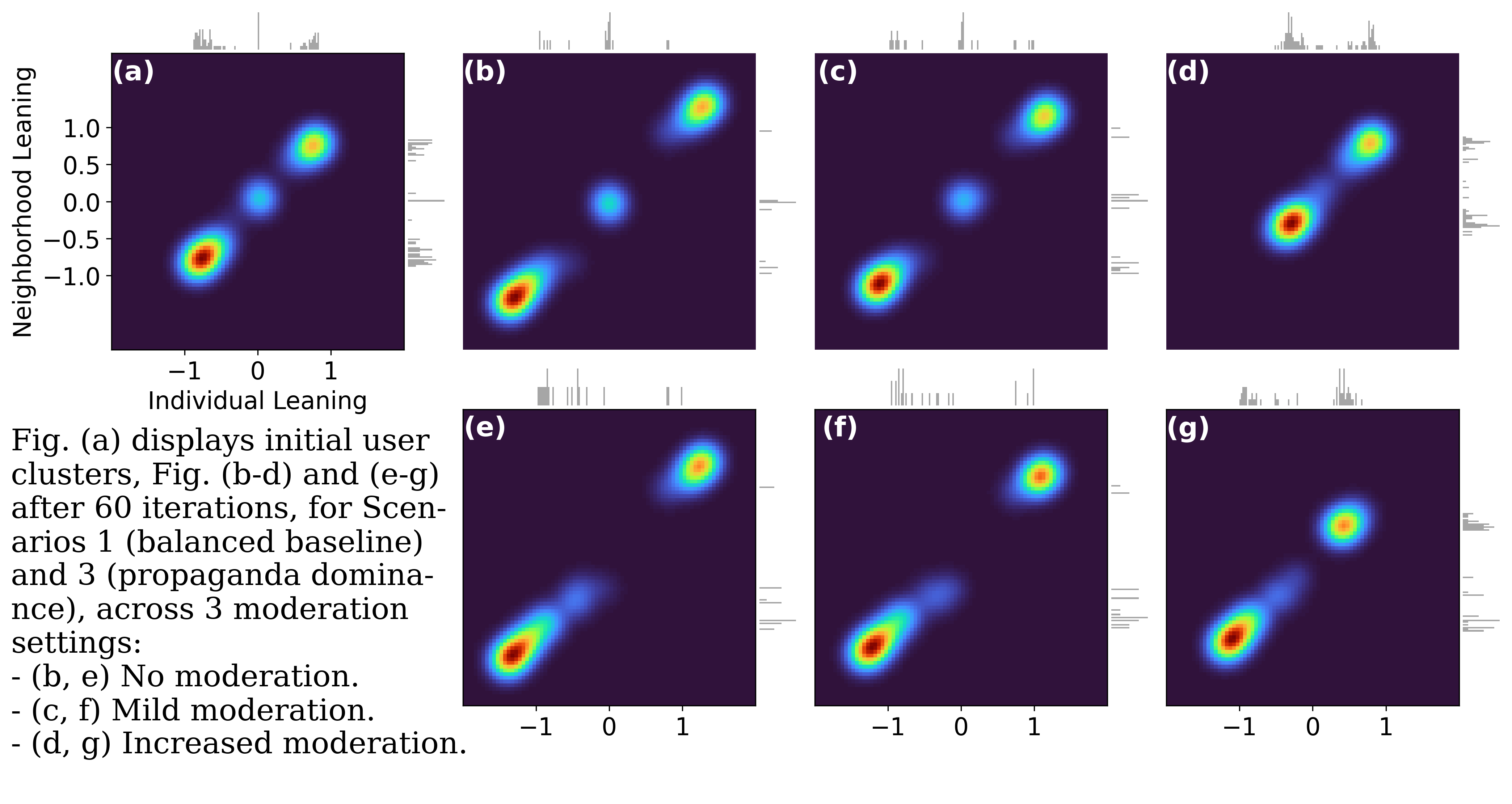}
    \caption{User opinion and neighborhood stance distribution at $t_0$ and $t_{60}$ without/with moderation of varied strengths.}
    \label{fig:rq3}
\end{figure}

\vspace{-5mm}
\subsection{\ref{rq3}: Impact on User Opinions and Stances}\label{sec:rq3_results}

Our analysis evaluates the JSD-* (read) alongside user opinion metrics---User Mean Stance (UMS) and User Mean Opinion Extremeness (UMOE), as detailed in Table~\ref{tab:rq3_9t}.

\emph{Changes in JSDs and User Metrics:}
Without moderation, JSDs typically rise, indicating increased polarization. 
Moderation through RR significantly lowers JSDs (read) but at the cost of substantial drops in CTR, as noted in Table~\ref{tab:rq2_9t}. RD and SD also reduce JSDs but with less severe impacts on CTRs. Both UMS and UMOE metrics increase without moderation, suggesting a trend toward more polarized views, especially in scenarios with extreme left-wing propaganda, which markedly shifts UMS to the left. 
While RR effectively neutralizes UMS with considerable CTR reductions, RD and SD manage to moderate UMS shifts more subtly. Conversely, KC often worsens UMS shifts, moving further from neutrality. 
All moderation approaches similarly reduce UMOE, indicating a general tempering of opinion extremeness across different methods.

\emph{Dynamic Observations of User Opinions:}
Enhancing the update coefficient $\gamma$ (Eq.~\ref{eq:user_update}) from 0.001 to 0.01 increases user sensitivity to recommendations, leading to more pronounced changes for visualization. Figure~\ref{fig:rq3} compares the start and end of the simulation under both balanced (S1) and extreme propaganda (S3) conditions. In the balanced scenario, user opinions without moderation become more segregated and extreme (\ref{fig:rq3}b). With moderation, opinions shift toward the center, decreasing extremeness (\ref{fig:rq3}c), and stronger moderation further centralizes user stances (\ref{fig:rq3}d). In the propaganda scenario, moderation limits the leftward drift of neutrality-leaning users and reduces overall extremity (\ref{fig:rq3}e, \ref{fig:rq3}f, and \ref{fig:rq3}g).

\emph{Answering \ref{rq3}}
The findings affirm the efficacy of cluster dispersal moderation strategies in reducing polarization within user opinions and stances across varied media landscapes. This underscores the potential of moderation techniques to balance exposure and subtly depolarize user opinions.

\emph{Caveat}
These results are contingent on specific simulation parameters and should be contextualized within the framework of our adopted user model and update dynamics. Alternative modeling approaches, such as those examined in \cite{chen2021opinion}, may yield different outcomes.
Anomalous behaviors observed in our simulated propaganda scenario, where JSD-O (read) decreases without moderation but increases under RR and KC, left for future work.

\vspace{-5mm}
\section{Conclusions}\label{sec:conclusion}
This study explored the concept of content-agnostic moderation to mitigate opinion polarization in personalized recommendation systems. The appeal of content-agnostic moderation is that it may achieve this goal without selectively suppressing particular (legal) views or opinions, based on the intrinsic properties of the items, which could be interpreted as censorship. We introduced and empirically validated two such moderation methods: \emph{Random Dispersal} and \emph{Similarity-based Dispersal}, relying solely on relational properties. 
Using a novel simulation environment, we empirically demonstrated that these methods show promise in enhancing stance neutrality and diversifying content exposure without relying on content details.

The main limitation or our work is that our observations and conclusions are contingent on the developed simulation environment and semi-synthetic data being sufficiently realistic.
We also acknowledge the potential for further exploration of the impact of user mental state modelling on the effectiveness of moderation strategies.
Nonetheless, we believe that our results at least serve as a proof of concept, provide valuable insights into the potential of content-agnostic moderation strategies in real-world recommendation systems, and open up new research directions in the field of recommendation systems and opinion dynamics.

\begin{acks}
\end{acks}

\bibliographystyle{ACM-Reference-Format}
\bibliography{ref}


\begin{thebibliography}{46}


\ifx \showCODEN    \undefined \def \showCODEN     #1{\unskip}     \fi
\ifx \showDOI      \undefined \def \showDOI       #1{#1}\fi
\ifx \showISBNx    \undefined \def \showISBNx     #1{\unskip}     \fi
\ifx \showISBNxiii \undefined \def \showISBNxiii  #1{\unskip}     \fi
\ifx \showISSN     \undefined \def \showISSN      #1{\unskip}     \fi
\ifx \showLCCN     \undefined \def \showLCCN      #1{\unskip}     \fi
\ifx \shownote     \undefined \def \shownote      #1{#1}          \fi
\ifx \showarticletitle \undefined \def \showarticletitle #1{#1}   \fi
\ifx \showURL      \undefined \def \showURL       {\relax}        \fi
\providecommand\bibfield[2]{#2}
\providecommand\bibinfo[2]{#2}
\providecommand\natexlab[1]{#1}
\providecommand\showeprint[2][]{arXiv:#2}

\bibitem[Abdollahpouri et~al\mbox{.}(2021)]%
        {abdollahpouri2021user}
\bibfield{author}{\bibinfo{person}{Himan Abdollahpouri},
  \bibinfo{person}{Masoud Mansoury}, \bibinfo{person}{Robin Burke},
  \bibinfo{person}{Bamshad Mobasher}, {and} \bibinfo{person}{Edward
  Malthouse}.} \bibinfo{year}{2021}\natexlab{}.
\newblock \showarticletitle{User-centered evaluation of popularity bias in
  recommender systems}. In \bibinfo{booktitle}{\emph{Proceedings of the 29th
  ACM conference on user modeling, adaptation and personalization}}.
  \bibinfo{pages}{119--129}.
\newblock


\bibitem[Anwar et~al\mbox{.}(2024)]%
        {anwar2024filter}
\bibfield{author}{\bibinfo{person}{Md~Sanzeed Anwar}, \bibinfo{person}{Grant
  Schoenebeck}, {and} \bibinfo{person}{Paramveer~S Dhillon}.}
  \bibinfo{year}{2024}\natexlab{}.
\newblock \showarticletitle{Filter Bubble or Homogenization? Disentangling the
  Long-Term Effects of Recommendations on User Consumption Patterns}.
\newblock \bibinfo{journal}{\emph{arXiv preprint arXiv:2402.15013}}
  (\bibinfo{year}{2024}).
\newblock


\bibitem[Badami(2017)]%
        {badami2017peeking}
\bibfield{author}{\bibinfo{person}{Mahsa Badami}.}
  \bibinfo{year}{2017}\natexlab{}.
\newblock \showarticletitle{Peeking into the other half of the glass: handling
  polarization in recommender systems.}
\newblock  (\bibinfo{year}{2017}).
\newblock


\bibitem[Badami et~al\mbox{.}(2017)]%
        {badami2017detecting}
\bibfield{author}{\bibinfo{person}{Mahsa Badami}, \bibinfo{person}{Olfa
  Nasraoui}, \bibinfo{person}{Welong Sun}, {and} \bibinfo{person}{Patrick
  Shafto}.} \bibinfo{year}{2017}\natexlab{}.
\newblock \showarticletitle{Detecting polarization in ratings: An automated
  pipeline and a preliminary quantification on several benchmark data sets}. In
  \bibinfo{booktitle}{\emph{2017 IEEE International Conference on Big Data (Big
  Data)}}. IEEE, \bibinfo{pages}{2682--2690}.
\newblock


\bibitem[Bakshy et~al\mbox{.}(2015)]%
        {bakshy2015exposure}
\bibfield{author}{\bibinfo{person}{Eytan Bakshy}, \bibinfo{person}{Solomon
  Messing}, {and} \bibinfo{person}{Lada~A Adamic}.}
  \bibinfo{year}{2015}\natexlab{}.
\newblock \showarticletitle{Exposure to ideologically diverse news and opinion
  on Facebook}.
\newblock \bibinfo{journal}{\emph{Science}} \bibinfo{volume}{348},
  \bibinfo{number}{6239} (\bibinfo{year}{2015}), \bibinfo{pages}{1130--1132}.
\newblock


\bibitem[Bolzern et~al\mbox{.}(2019)]%
        {bolzern2019opinion}
\bibfield{author}{\bibinfo{person}{Paolo Bolzern}, \bibinfo{person}{Patrizio
  Colaneri}, {and} \bibinfo{person}{Giuseppe De~Nicolao}.}
  \bibinfo{year}{2019}\natexlab{}.
\newblock \showarticletitle{Opinion influence and evolution in social networks:
  A Markovian agents model}.
\newblock \bibinfo{journal}{\emph{Automatica}}  \bibinfo{volume}{100}
  (\bibinfo{year}{2019}), \bibinfo{pages}{219--230}.
\newblock


\bibitem[Bressan et~al\mbox{.}(2016)]%
        {bressan2016limits}
\bibfield{author}{\bibinfo{person}{Marco Bressan}, \bibinfo{person}{Stefano
  Leucci}, \bibinfo{person}{Alessandro Panconesi}, \bibinfo{person}{Prabhakar
  Raghavan}, {and} \bibinfo{person}{Erisa Terolli}.}
  \bibinfo{year}{2016}\natexlab{}.
\newblock \showarticletitle{The limits of popularity-based recommendations, and
  the role of social ties}. In \bibinfo{booktitle}{\emph{Proceedings of the
  22nd ACM SIGKDD International Conference on Knowledge Discovery and Data
  Mining}}. \bibinfo{pages}{745--754}.
\newblock


\bibitem[Chen et~al\mbox{.}(2021)]%
        {chen2021opinion}
\bibfield{author}{\bibinfo{person}{Xi Chen}, \bibinfo{person}{Panayiotis
  Tsaparas}, \bibinfo{person}{Jefrey Lijffijt}, {and} \bibinfo{person}{Tijl
  De~Bie}.} \bibinfo{year}{2021}\natexlab{}.
\newblock \showarticletitle{Opinion dynamics with backfire effect and biased
  assimilation}.
\newblock \bibinfo{journal}{\emph{PloS one}} \bibinfo{volume}{16},
  \bibinfo{number}{9} (\bibinfo{year}{2021}), \bibinfo{pages}{e0256922}.
\newblock


\bibitem[Chitra and Musco(2020)]%
        {chitra2020analyzing}
\bibfield{author}{\bibinfo{person}{Uthsav Chitra} {and}
  \bibinfo{person}{Christopher Musco}.} \bibinfo{year}{2020}\natexlab{}.
\newblock \showarticletitle{Analyzing the impact of filter bubbles on social
  network polarization}. In \bibinfo{booktitle}{\emph{Proceedings of the 13th
  International Conference on Web Search and Data Mining}}.
  \bibinfo{pages}{115--123}.
\newblock


\bibitem[Chuai et~al\mbox{.}(2023)]%
        {chuai2023roll}
\bibfield{author}{\bibinfo{person}{Yuwei Chuai}, \bibinfo{person}{Haoye Tian},
  \bibinfo{person}{Nicolas Pr{\"o}llochs}, {and} \bibinfo{person}{Gabriele
  Lenzini}.} \bibinfo{year}{2023}\natexlab{}.
\newblock \showarticletitle{The Roll-Out of Community Notes Did Not Reduce
  Engagement With Misinformation on Twitter}.
\newblock \bibinfo{journal}{\emph{arXiv preprint arXiv:2307.07960}}
  (\bibinfo{year}{2023}).
\newblock


\bibitem[Dandekar et~al\mbox{.}(2013)]%
        {dandekar2013biased}
\bibfield{author}{\bibinfo{person}{Pranav Dandekar}, \bibinfo{person}{Ashish
  Goel}, {and} \bibinfo{person}{David~T Lee}.} \bibinfo{year}{2013}\natexlab{}.
\newblock \showarticletitle{Biased assimilation, homophily, and the dynamics of
  polarization}.
\newblock \bibinfo{journal}{\emph{Proceedings of the National Academy of
  Sciences}} \bibinfo{volume}{110}, \bibinfo{number}{15}
  (\bibinfo{year}{2013}), \bibinfo{pages}{5791--5796}.
\newblock


\bibitem[Datar et~al\mbox{.}(2004)]%
        {datar2004locality}
\bibfield{author}{\bibinfo{person}{Mayur Datar}, \bibinfo{person}{Nicole
  Immorlica}, \bibinfo{person}{Piotr Indyk}, {and} \bibinfo{person}{Vahab~S
  Mirrokni}.} \bibinfo{year}{2004}\natexlab{}.
\newblock \showarticletitle{Locality-sensitive hashing scheme based on p-stable
  distributions}. In \bibinfo{booktitle}{\emph{Proceedings of the twentieth
  annual symposium on Computational geometry}}. \bibinfo{pages}{253--262}.
\newblock


\bibitem[Davis and Rosen(2019)]%
        {davis2019open}
\bibfield{author}{\bibinfo{person}{Antigone Davis} {and} \bibinfo{person}{Guy
  Rosen}.} \bibinfo{year}{2019}\natexlab{}.
\newblock \showarticletitle{Open-sourcing photo-and video-matching technology
  to make the internet safer}.
\newblock \bibinfo{journal}{\emph{Facebook Newsroom}} (\bibinfo{year}{2019}).
\newblock


\bibitem[Doherty et~al\mbox{.}(2021)]%
        {doherty2021beyond}
\bibfield{author}{\bibinfo{person}{Carroll Doherty}, \bibinfo{person}{Jocelyn
  Kiley}, \bibinfo{person}{Nida Asheer}, {and} \bibinfo{person}{Calvin
  Jordan}.} \bibinfo{year}{2021}\natexlab{}.
\newblock \showarticletitle{Beyond red vs. blue: The political typology}.
\newblock \bibinfo{journal}{\emph{Pew Research Center}} (\bibinfo{year}{2021}).
\newblock


\bibitem[{European Commission}(2020)]%
        {eu2020digital}
\bibfield{author}{\bibinfo{person}{{European Commission}}.}
  \bibinfo{year}{2020}\natexlab{}.
\newblock \bibinfo{title}{Proposal for a Regulation of the European Parliament
  and of the Council on the Single Market for Digital Services (Digital
  Services Act) and amending Directive}.
\newblock
  \bibinfo{howpublished}{\url{https://eur-lex.europa.eu/EN/legal-content/summary/digital-services-act.html}}.
\newblock
\newblock
\shownote{Accessed: 2024-04-15}.


\bibitem[Flaxman et~al\mbox{.}(2016)]%
        {flaxman2016filter}
\bibfield{author}{\bibinfo{person}{Seth Flaxman}, \bibinfo{person}{Sharad
  Goel}, {and} \bibinfo{person}{Justin~M Rao}.}
  \bibinfo{year}{2016}\natexlab{}.
\newblock \showarticletitle{Filter bubbles, echo chambers, and online news
  consumption}.
\newblock \bibinfo{journal}{\emph{Public opinion quarterly}}
  \bibinfo{volume}{80}, \bibinfo{number}{S1} (\bibinfo{year}{2016}),
  \bibinfo{pages}{298--320}.
\newblock


\bibitem[Geschke et~al\mbox{.}(2019)]%
        {geschke2019triple}
\bibfield{author}{\bibinfo{person}{Daniel Geschke}, \bibinfo{person}{Jan
  Lorenz}, {and} \bibinfo{person}{Peter Holtz}.}
  \bibinfo{year}{2019}\natexlab{}.
\newblock \showarticletitle{The triple-filter bubble: Using agent-based
  modelling to test a meta-theoretical framework for the emergence of filter
  bubbles and echo chambers}.
\newblock \bibinfo{journal}{\emph{British Journal of Social Psychology}}
  \bibinfo{volume}{58}, \bibinfo{number}{1} (\bibinfo{year}{2019}),
  \bibinfo{pages}{129--149}.
\newblock


\bibitem[Gillespie(2022)]%
        {gillespie2022not}
\bibfield{author}{\bibinfo{person}{Tarleton Gillespie}.}
  \bibinfo{year}{2022}\natexlab{}.
\newblock \showarticletitle{Do not recommend? Reduction as a form of content
  moderation}.
\newblock \bibinfo{journal}{\emph{Social Media+ Society}} \bibinfo{volume}{8},
  \bibinfo{number}{3} (\bibinfo{year}{2022}),
  \bibinfo{pages}{20563051221117552}.
\newblock


\bibitem[Gongane et~al\mbox{.}(2022)]%
        {gongane2022detection}
\bibfield{author}{\bibinfo{person}{Vaishali~U Gongane},
  \bibinfo{person}{Mousami~V Munot}, {and} \bibinfo{person}{Alwin~D Anuse}.}
  \bibinfo{year}{2022}\natexlab{}.
\newblock \showarticletitle{Detection and moderation of detrimental content on
  social media platforms: Current status and future directions}.
\newblock \bibinfo{journal}{\emph{Social Network Analysis and Mining}}
  \bibinfo{volume}{12}, \bibinfo{number}{1} (\bibinfo{year}{2022}),
  \bibinfo{pages}{129}.
\newblock


\bibitem[Gorwa et~al\mbox{.}(2020)]%
        {gorwa2020algorithmic}
\bibfield{author}{\bibinfo{person}{Robert Gorwa}, \bibinfo{person}{Reuben
  Binns}, {and} \bibinfo{person}{Christian Katzenbach}.}
  \bibinfo{year}{2020}\natexlab{}.
\newblock \showarticletitle{Algorithmic content moderation: Technical and
  political challenges in the automation of platform governance}.
\newblock \bibinfo{journal}{\emph{Big Data \& Society}} \bibinfo{volume}{7},
  \bibinfo{number}{1} (\bibinfo{year}{2020}),
  \bibinfo{pages}{2053951719897945}.
\newblock


\bibitem[Grimmelmann(2015)]%
        {grimmelmann2015virtues}
\bibfield{author}{\bibinfo{person}{James Grimmelmann}.}
  \bibinfo{year}{2015}\natexlab{}.
\newblock \showarticletitle{The virtues of moderation}.
\newblock \bibinfo{journal}{\emph{Yale JL \& Tech.}}  \bibinfo{volume}{17}
  (\bibinfo{year}{2015}), \bibinfo{pages}{42}.
\newblock


\bibitem[Heitz et~al\mbox{.}(2022)]%
        {heitz2022benefits}
\bibfield{author}{\bibinfo{person}{Lucien Heitz}, \bibinfo{person}{Juliane~A
  Lischka}, \bibinfo{person}{Alena Birrer}, \bibinfo{person}{Bibek Paudel},
  \bibinfo{person}{Suzanne Tolmeijer}, \bibinfo{person}{Laura Laugwitz}, {and}
  \bibinfo{person}{Abraham Bernstein}.} \bibinfo{year}{2022}\natexlab{}.
\newblock \showarticletitle{Benefits of diverse news recommendations for
  democracy: A user study}.
\newblock \bibinfo{journal}{\emph{Digital Journalism}} \bibinfo{volume}{10},
  \bibinfo{number}{10} (\bibinfo{year}{2022}), \bibinfo{pages}{1710--1730}.
\newblock


\bibitem[Jambor and Wang(2010)]%
        {jambor2010optimizing}
\bibfield{author}{\bibinfo{person}{Tamas Jambor} {and} \bibinfo{person}{Jun
  Wang}.} \bibinfo{year}{2010}\natexlab{}.
\newblock \showarticletitle{Optimizing Multiple Objectives in Collaborative
  Filtering}. In \bibinfo{booktitle}{\emph{Proceedings of the Fourth {{ACM}}
  Conference on {{Recommender}} Systems}}. \bibinfo{publisher}{ACM},
  \bibinfo{address}{Barcelona Spain}, \bibinfo{pages}{55--62}.
\newblock
\showISBNx{978-1-60558-906-0}
\urldef\tempurl%
\url{https://doi.org/10.1145/1864708.1864723}
\showDOI{\tempurl}


\bibitem[Jiang et~al\mbox{.}(2019)]%
        {jiang2019degenerate}
\bibfield{author}{\bibinfo{person}{Ray Jiang}, \bibinfo{person}{Silvia
  Chiappa}, \bibinfo{person}{Tor Lattimore}, \bibinfo{person}{Andr{\'a}s
  Gy{\"o}rgy}, {and} \bibinfo{person}{Pushmeet Kohli}.}
  \bibinfo{year}{2019}\natexlab{}.
\newblock \showarticletitle{Degenerate feedback loops in recommender systems}.
  In \bibinfo{booktitle}{\emph{Proceedings of the 2019 AAAI/ACM Conference on
  AI, Ethics, and Society}}. \bibinfo{pages}{383--390}.
\newblock


\bibitem[Jugovac et~al\mbox{.}(2017)]%
        {jugovac2017efficient}
\bibfield{author}{\bibinfo{person}{Michael Jugovac}, \bibinfo{person}{Dietmar
  Jannach}, {and} \bibinfo{person}{Lukas Lerche}.}
  \bibinfo{year}{2017}\natexlab{}.
\newblock \showarticletitle{Efficient optimization of multiple recommendation
  quality factors according to individual user tendencies}.
\newblock \bibinfo{journal}{\emph{Expert Systems with Applications}}
  \bibinfo{volume}{81} (\bibinfo{year}{2017}), \bibinfo{pages}{321--331}.
\newblock


\bibitem[Knudsen(2023)]%
        {knudsen2023modeling}
\bibfield{author}{\bibinfo{person}{Erik Knudsen}.}
  \bibinfo{year}{2023}\natexlab{}.
\newblock \showarticletitle{Modeling news recommender systems’ conditional
  effects on selective exposure: evidence from two online experiments}.
\newblock \bibinfo{journal}{\emph{Journal of Communication}}
  \bibinfo{volume}{73}, \bibinfo{number}{2} (\bibinfo{year}{2023}),
  \bibinfo{pages}{138--149}.
\newblock


\bibitem[Lin(1991)]%
        {lin1991divergence}
\bibfield{author}{\bibinfo{person}{Jianhua Lin}.}
  \bibinfo{year}{1991}\natexlab{}.
\newblock \showarticletitle{Divergence measures based on the Shannon entropy}.
\newblock \bibinfo{journal}{\emph{IEEE Transactions on Information theory}}
  \bibinfo{volume}{37}, \bibinfo{number}{1} (\bibinfo{year}{1991}),
  \bibinfo{pages}{145--151}.
\newblock


\bibitem[Liu et~al\mbox{.}(2023)]%
        {liu2023topic}
\bibfield{author}{\bibinfo{person}{Dairui Liu}, \bibinfo{person}{Derek Greene},
  \bibinfo{person}{Irene Li}, {and} \bibinfo{person}{Ruihai Dong}.}
  \bibinfo{year}{2023}\natexlab{}.
\newblock \showarticletitle{Topic-Centric Explanations for News
  Recommendation}.
\newblock \bibinfo{journal}{\emph{arXiv preprint arXiv:2306.07506}}
  (\bibinfo{year}{2023}).
\newblock


\bibitem[Liu et~al\mbox{.}(2021)]%
        {liu2021interaction}
\bibfield{author}{\bibinfo{person}{Ping Liu}, \bibinfo{person}{Karthik
  Shivaram}, \bibinfo{person}{Aron Culotta}, \bibinfo{person}{Matthew~A
  Shapiro}, {and} \bibinfo{person}{Mustafa Bilgic}.}
  \bibinfo{year}{2021}\natexlab{}.
\newblock \showarticletitle{The interaction between political typology and
  filter bubbles in news recommendation algorithms}. In
  \bibinfo{booktitle}{\emph{Proceedings of the Web Conference 2021}}.
  \bibinfo{pages}{3791--3801}.
\newblock


\bibitem[Narayanan(2023)]%
        {narayanan2023understanding}
\bibfield{author}{\bibinfo{person}{Arvind Narayanan}.}
  \bibinfo{year}{2023}\natexlab{}.
\newblock \showarticletitle{Understanding social media recommendation
  algorithms}.
\newblock  (\bibinfo{year}{2023}).
\newblock


\bibitem[Nguyen(2020)]%
        {nguyen2020echo}
\bibfield{author}{\bibinfo{person}{C~Thi Nguyen}.}
  \bibinfo{year}{2020}\natexlab{}.
\newblock \showarticletitle{Echo chambers and epistemic bubbles}.
\newblock \bibinfo{journal}{\emph{Episteme}} \bibinfo{volume}{17},
  \bibinfo{number}{2} (\bibinfo{year}{2020}), \bibinfo{pages}{141--161}.
\newblock


\bibitem[Nguyen et~al\mbox{.}(2014)]%
        {nguyen2014exploring}
\bibfield{author}{\bibinfo{person}{Tien~T Nguyen}, \bibinfo{person}{Pik-Mai
  Hui}, \bibinfo{person}{F~Maxwell Harper}, \bibinfo{person}{Loren Terveen},
  {and} \bibinfo{person}{Joseph~A Konstan}.} \bibinfo{year}{2014}\natexlab{}.
\newblock \showarticletitle{Exploring the filter bubble: the effect of using
  recommender systems on content diversity}. In
  \bibinfo{booktitle}{\emph{Proceedings of the 23rd international conference on
  World wide web}}. \bibinfo{pages}{677--686}.
\newblock


\bibitem[Pansanella et~al\mbox{.}(2023)]%
        {pansanella2023mass}
\bibfield{author}{\bibinfo{person}{Valentina Pansanella},
  \bibinfo{person}{Alina S{\^\i}rbu}, \bibinfo{person}{Janos Kertesz}, {and}
  \bibinfo{person}{Giulio Rossetti}.} \bibinfo{year}{2023}\natexlab{}.
\newblock \showarticletitle{Mass media impact on opinion evolution in biased
  digital environments: a bounded confidence model}.
\newblock \bibinfo{journal}{\emph{Scientific Reports}} \bibinfo{volume}{13},
  \bibinfo{number}{1} (\bibinfo{year}{2023}), \bibinfo{pages}{14600}.
\newblock


\bibitem[Patankar et~al\mbox{.}(2019)]%
        {patankar2019bias}
\bibfield{author}{\bibinfo{person}{Anish Patankar}, \bibinfo{person}{Joy Bose},
  {and} \bibinfo{person}{Harshit Khanna}.} \bibinfo{year}{2019}\natexlab{}.
\newblock \showarticletitle{A bias aware news recommendation system}. In
  \bibinfo{booktitle}{\emph{2019 IEEE 13th International Conference on Semantic
  Computing (ICSC)}}. IEEE, \bibinfo{pages}{232--238}.
\newblock


\bibitem[Patro et~al\mbox{.}(2020)]%
        {patro2020fairrec}
\bibfield{author}{\bibinfo{person}{Gourab~K Patro}, \bibinfo{person}{Arpita
  Biswas}, \bibinfo{person}{Niloy Ganguly}, \bibinfo{person}{Krishna~P
  Gummadi}, {and} \bibinfo{person}{Abhijnan Chakraborty}.}
  \bibinfo{year}{2020}\natexlab{}.
\newblock \showarticletitle{Fairrec: Two-sided fairness for personalized
  recommendations in two-sided platforms}. In
  \bibinfo{booktitle}{\emph{Proceedings of the web conference 2020}}.
  \bibinfo{pages}{1194--1204}.
\newblock


\bibitem[Perra and Rocha(2019)]%
        {perra2019modelling}
\bibfield{author}{\bibinfo{person}{Nicola Perra} {and} \bibinfo{person}{Luis~EC
  Rocha}.} \bibinfo{year}{2019}\natexlab{}.
\newblock \showarticletitle{Modelling opinion dynamics in the age of
  algorithmic personalisation}.
\newblock \bibinfo{journal}{\emph{Scientific reports}} \bibinfo{volume}{9},
  \bibinfo{number}{1} (\bibinfo{year}{2019}), \bibinfo{pages}{7261}.
\newblock


\bibitem[Ribeiro et~al\mbox{.}(2012)]%
        {ribeiro2012paretoefficient}
\bibfield{author}{\bibinfo{person}{Marco~Tulio Ribeiro},
  \bibinfo{person}{Anisio Lacerda}, \bibinfo{person}{Adriano Veloso}, {and}
  \bibinfo{person}{Nivio Ziviani}.} \bibinfo{year}{2012}\natexlab{}.
\newblock \showarticletitle{Pareto-Efficient Hybridization for Multi-Objective
  Recommender Systems}. In \bibinfo{booktitle}{\emph{Proceedings of the Sixth
  {{ACM}} Conference on {{Recommender}} Systems}}. \bibinfo{publisher}{ACM},
  \bibinfo{address}{Dublin Ireland}, \bibinfo{pages}{19--26}.
\newblock
\showISBNx{978-1-4503-1270-7}
\urldef\tempurl%
\url{https://doi.org/10.1145/2365952.2365962}
\showDOI{\tempurl}


\bibitem[Rossi et~al\mbox{.}(2021)]%
        {rossi2021closed}
\bibfield{author}{\bibinfo{person}{Wilbert~Samuel Rossi},
  \bibinfo{person}{Jan~Willem Polderman}, {and} \bibinfo{person}{Paolo
  Frasca}.} \bibinfo{year}{2021}\natexlab{}.
\newblock \showarticletitle{The closed loop between opinion formation and
  personalized recommendations}.
\newblock \bibinfo{journal}{\emph{IEEE Transactions on Control of Network
  Systems}} \bibinfo{volume}{9}, \bibinfo{number}{3} (\bibinfo{year}{2021}),
  \bibinfo{pages}{1092--1103}.
\newblock


\bibitem[Schmidt and Wiegand(2017)]%
        {schmidt2017survey}
\bibfield{author}{\bibinfo{person}{Anna Schmidt} {and} \bibinfo{person}{Michael
  Wiegand}.} \bibinfo{year}{2017}\natexlab{}.
\newblock \showarticletitle{A survey on hate speech detection using natural
  language processing}. In \bibinfo{booktitle}{\emph{Proceedings of the fifth
  international workshop on natural language processing for social media}}.
  \bibinfo{pages}{1--10}.
\newblock


\bibitem[SEYMEN et~al\mbox{.}(2021)]%
        {seymenunified2021}
\bibfield{author}{\bibinfo{person}{SINAN SEYMEN}, \bibinfo{person}{HIMAN
  ABDOLLAHPOURI}, {and} \bibinfo{person}{EDWARD~C MALTHOUSE}.}
  \bibinfo{year}{2021}\natexlab{}.
\newblock \showarticletitle{A unified optimization toolbox for solving
  popularity bias, fairness, and diversity in recommender systems}.
\newblock


\bibitem[Spinelli and Crovella(2017)]%
        {spinelli2017closed}
\bibfield{author}{\bibinfo{person}{Larissa Spinelli} {and}
  \bibinfo{person}{Mark Crovella}.} \bibinfo{year}{2017}\natexlab{}.
\newblock \showarticletitle{Closed-loop opinion formation}. In
  \bibinfo{booktitle}{\emph{Proceedings of the 2017 ACM on Web Science
  Conference}}. \bibinfo{pages}{73--82}.
\newblock


\bibitem[Stray(2021)]%
        {stray2021designing}
\bibfield{author}{\bibinfo{person}{Jonathan Stray}.}
  \bibinfo{year}{2021}\natexlab{}.
\newblock \showarticletitle{Designing recommender systems to depolarize}.
\newblock \bibinfo{journal}{\emph{arXiv preprint arXiv:2107.04953}}
  (\bibinfo{year}{2021}).
\newblock


\bibitem[Turillazzi et~al\mbox{.}(2023)]%
        {turillazzi2023digital}
\bibfield{author}{\bibinfo{person}{Aina Turillazzi},
  \bibinfo{person}{Mariarosaria Taddeo}, \bibinfo{person}{Luciano Floridi},
  {and} \bibinfo{person}{Federico Casolari}.} \bibinfo{year}{2023}\natexlab{}.
\newblock \showarticletitle{The digital services act: an analysis of its
  ethical, legal, and social implications}.
\newblock \bibinfo{journal}{\emph{Law, Innovation and Technology}}
  \bibinfo{volume}{15}, \bibinfo{number}{1} (\bibinfo{year}{2023}),
  \bibinfo{pages}{83--106}.
\newblock


\bibitem[Wu et~al\mbox{.}(2021)]%
        {wu2021tfrom}
\bibfield{author}{\bibinfo{person}{Y. Wu}, \bibinfo{person}{J. Cao},
  \bibinfo{person}{G. Xu}, {and} \bibinfo{person}{Y. Tan}.}
  \bibinfo{year}{2021}\natexlab{}.
\newblock \showarticletitle{{{TFROM}}: {{A Two-sidied Fairness-Aware
  Recommendation Modeleldel}} for {{Both Customers}} and {{Providers}}}. In
  \bibinfo{booktitle}{\emph{{{SIGIR}}'21: {{The}} 44th {{International ACM
  SIGIR Conference}} on {{Research}} and {{Development}} in {{Information
  Retrieval}}}}. \bibinfo{publisher}{{ACM}}.
\newblock


\bibitem[Zhang et~al\mbox{.}(2023)]%
        {zhang2023evolution}
\bibfield{author}{\bibinfo{person}{Han Zhang}, \bibinfo{person}{Ziwei Zhu},
  {and} \bibinfo{person}{James Caverlee}.} \bibinfo{year}{2023}\natexlab{}.
\newblock \showarticletitle{Evolution of filter bubbles and polarization in
  news recommendation}. In \bibinfo{booktitle}{\emph{European Conference on
  Information Retrieval}}. Springer, \bibinfo{pages}{685--693}.
\newblock


\bibitem[Zhu et~al\mbox{.}(2021)]%
        {zhu2021popularity}
\bibfield{author}{\bibinfo{person}{Ziwei Zhu}, \bibinfo{person}{Yun He},
  \bibinfo{person}{Xing Zhao}, \bibinfo{person}{Yin Zhang},
  \bibinfo{person}{Jianling Wang}, {and} \bibinfo{person}{James Caverlee}.}
  \bibinfo{year}{2021}\natexlab{}.
\newblock \showarticletitle{Popularity-opportunity bias in collaborative
  filtering}. In \bibinfo{booktitle}{\emph{Proceedings of the 14th ACM
  International Conference on Web Search and Data Mining}}.
  \bibinfo{pages}{85--93}.
\newblock


\end{thebibliography}

\newpage
\appendix
\section{Detailed Proofs}\label{app:proofs}

This appendix provides formal proofs for the propositions outlined in the theoretical analysis section.

\subsection{Proof of Theorem from Proposition 1}

\textbf{Claim:} A content-agnostic function \(f\) cannot guarantee a targeted stance distribution across all recommendations because it cannot distinguish \(\pi_1\) from \(\pi_2\) based solely on content-agnostic features.

\begin{proof}
Assume for contradiction that there exists a content-agnostic function \( f: \Pi \rightarrow \Pi \) capable of modifying any given recommendation configuration \(\pi\) to achieve a targeted distribution \(D\).

Consider two recommendation configurations:
\begin{enumerate}
    \item \(\pi_1\) consisting entirely of items with a predominant stance \(s_1\).
    \item \(\pi_2\) where items are evenly distributed among all stances in \(S\).
\end{enumerate}
Both \(\pi_1\) and \(\pi_2\) are potential inputs to \(f\), but without stance information, \(f\) treats them identically.

To achieve \(D\), \(f(\pi_1)\) would need to substantially alter its item composition towards a more balanced distribution, akin to \(f(\pi_2)\). However, since \(f\) is content-agnostic and cannot access or differentiate based on item stances, it lacks the necessary information to make such distinctions or adjustments.

Thus, it is impossible for \(f\) to consistently transform \(\pi_1\) into a configuration that aligns with \(D\) while treating \(\pi_2\) identically. This contradiction implies that no such function \(f\) can exist that meets the criteria for all \(\pi \in \Pi\).
\end{proof}

\subsection{Proof of Theorem from Proposition 2}

\textbf{Claim:} If \(f\) were capable of producing an output \(\pi'\) that conforms to \(D\), consistently training \(f\) to realize \(\pi'\) using a content-agnostic learning approach is unfeasible, as the algorithm cannot differentiate \(\pi'\) from other outputs \(\pi''\) that do not meet \(D\).

\begin{proof}
Assume hypothetically that \(f\) could produce \(\pi'\) with the desired distribution \(D\) from some input \(\pi\). Consider an alternative output \(\pi''\), which does not conform to \(D\), potentially generated by \(f\) from the same or a different input.

During training, the lack of stance information means that any learning algorithm or loss function applied to optimize \(f\) cannot distinguish effectively between \(\pi'\) and \(\pi''\) based on their adherence to \(D\).

This indistinguishability leads to a training process where \(\pi'\) and \(\pi''\) are equally likely outcomes of \(f\). Without a reliable method to favor \(\pi'\) over \(\pi''\), the algorithm cannot consistently produce the desired output, resulting in unpredictable and unreliable moderation outcomes.

Therefore, it is theoretically impossible to train or design a function \(f\) that can reliably produce outputs aligning with \(D\) without a mechanism to evaluate or differentiate based on the stance distribution of its outputs.
\end{proof}

\subsection{Proof that Egalitarian Exposure Leads to Uniform Distribution of Stances}
Assume that the set of items \(\mathcal{I}\) in the recommendation system is uniformly distributed across a set of stances \(S\). That is, each stance \(s \in S\) is equally likely to appear among the items in \(\mathcal{I}\).

\textbf{Proposition:} Uniform exposure of each item in \(\mathcal{I}\) results in a uniform distribution of stances across all recommendations.

\begin{proof}
Let \(|\mathcal{I}_s|\) represent the number of items in \(\mathcal{I}\) associated with stance \(s\). Since the distribution of stances in \(\mathcal{I}\) is uniform, we have:
    \[
    |\mathcal{I}_s| = \frac{|\mathcal{I}|}{|S|} \quad \text{for each } s \in S.
    \]
Given that each item \(i \in \mathcal{I}\) is exposed equally, let \(e_i\) denote the uniform exposure frequency for each item. Therefore, the exposure frequency of items associated with any stance \(s\) is:
    \[
    E_s = \sum_{i \in \mathcal{I}_s} e_i = e_i \times |\mathcal{I}_s| = e_i \times \frac{|\mathcal{I}|}{|S|}
    \]
Since \(E_s\) is proportional to \(|\mathcal{I}_s|\) and every \(|\mathcal{I}_s|\) is equal due to the uniform distribution of stances within \(\mathcal{I}\), \(E_s\) is identical for each stance \(s\). Hence, the overall exposure of stances across all recommendations is also uniform.
\end{proof}

\begin{algorithm}[!ht]
    \small
    \SetAlgoLined
    \KwData{The recommendation matrix $RM_t$, the aggregated recommendation matrix $RM_{agg, t} \in \mathbb{N}^{m \times n}$, minimal initial allocation limit $\alpha_{\text{min}} = 1$, top\_k}
    \KwResult{Adjusted recommendation matrix $X$}
    $m \gets \text{number of users}$\;
    $n \gets \text{number of items}$\;
    \For{$j \in \{1, \ldots, n\}$}{
        $q_j \gets \left\lceil \frac{\alpha_{\text{min}} \cdot m \cdot \text{top\_k}}{n} \right\rceil$ \tcp*[r]{initial copies per item}
    }
    $total\_needed \gets m \cdot \text{top\_k}$\;
    $total\_copies \gets \sum_{j=1}^n q_j$\;
    \While{$total\_copies < total\_needed$}{
        $\alpha_{\text{min}} \gets \alpha_{\text{min}} + 0.01$\;
        \For{$j \in \{1, \ldots, n\}$}{
            $q_j \gets \left\lceil \frac{\alpha_{\text{min}} \cdot m \cdot \text{top\_k}}{n} \right\rceil$ \tcp*[r]{dynamically adjust copies per item}
        }
        $total\_copies \gets \sum_{j=1}^n q_j$\;
    }
    \For{$i \in \{1, \ldots, m\}$}{
        \For{$j \in \{1, \ldots, n\}$}{
            $X_{ij} \gets$ Choose $RM_{t, ij}$ if not exceeding $q_j - RM_{agg, ij}$
            \tcp*[r]{choose the item with top rank if possible}
        }
    }
    \caption{RoundRobin with Dynamic Quota Adjustment}
    \label{alg:rr}
\end{algorithm}

\section{Details of Simulation Settings}\label{app:experiment_settings}
This section outlines the primary differences between our simulation setup and existing models, enhancing the realism and specificity of our experimental approach:

\begin{itemize}
    \item \textbf{Oracle Recommender Dynamics:} Unlike typical simulations that calculate expectations in a single step, our Oracle recommender operates over time, making recommendations in each of \( t \times k \) iterations for each of the 5000 runs. This dynamic approach helps investigate user behavior changes over time and allows for the incorporation of user preference updates. 
    
    \item \textbf{Initial Interaction Matrix Bootstrapping:} To minimize the impact of cold items, each user is initially exposed to 50 items. Subsequently, we identify and expose each cold item to 10 additional users to ensure that all items receive some initial interaction, thus eliminating early bias in item exposure.
    
    \item \textbf{Multiple Recommendations (\( k > 1 \)):} We set \( k \) to a value greater than one to mimic more realistic user interactions and provide flexibility in testing post-processing moderation techniques, such as RR, which are designed to handle multiple recommendations per session.    
  
    \item \textbf{Controlled Item Sampling:} We sample different proportions of items from each stance to facilitate comparisons of moderation effects across various scenarios.
    
    \item \textbf{Simplification of Topics:} For simplicity and clarity, we assume all items and interactions pertain to a single topic, allowing us to focus on stance dynamics without the complication of multiple thematic areas.
    
    \item \textbf{User Model Variations:} We explore both static and dynamic user models—updating and non-updating—to differentiate the effects of users' self-amplification of preferences from the inherent bias introduced by the recommender system. This separation helps in understanding the endogenous and exogenous factors affecting user behavior.
\end{itemize}

These adjustments and settings are designed to provide a robust framework for assessing the impacts of different recommender systems and moderation strategies, ensuring that results reflect variations in system behavior and user interactions under controlled yet realistic conditions.

\section{Details of the moderation methods}\label{app:moderators}
\subsection{Knapsack Constrained Optimization (KC)}\label{app:kc}

\emph{Model Setup}
Let \( m \) and \( n \) represent the number of users and items, respectively. The recommendation matrix output by a recommender at $t_0$ \( RM_t \in \{0,1\}^{m \times n} \) with entries \(RM_{t, ij}\) indicating whether item \( j \) is recommended to user \( i \). Item popularity weights are calculated as:
\[ w_j = \frac{\sum_{i=1}^{m} RM_{t, ij}}{\sum_{i=1}^{m} \sum_{j=1}^{n} RM_{t, ij}} \]

The popularity constraint is set by \( \alpha \), defined as:
\[ \alpha = \left(\sum_{i=1}^{m} \sum_{j=1}^{n} RM_{t, ij} w_j \right) \lambda \]
where \( \lambda \) is a hyperparameter that adjusts the tightness of the constraint.

\emph{Optimization Problem}
Define binary variables \( X_{ij} \) corresponding to the elements of the modified recommendation matrix \( X \), where \( X_{ij} = 1 \) if item \( j \) is recommended to user \( i \), and \( X_{ij} = 0 \) otherwise.

The objective is to minimize the absolute difference between the original and modified recommendations:
\[ \min \sum_{i=1}^{m} \sum_{j=1}^{n} |RM_{t, ij} - X_{ij}| \]

\emph{Constraints}
\begin{itemize}
    \item \textbf{Popularity Constraint:} The aggregated popularity of the recommended items should not exceed \( \alpha \):
    \[ \sum_{i=1}^{m} \sum_{j=1}^{n} X_{ij} w_j \leq \alpha \]
    \item \textbf{Row Sum Constraint:} Each user \( i \) should receive exactly \( k \) recommendations:
    \[ \sum_{j=1}^{n} X_{ij} = k \quad \text{for all } i \]
\end{itemize}

\subsection{RoundRobin Allocation with Dynamic Quota Adjustment (RR)}\label{app:rr}

\emph{Algorithm Description}
This method dynamically adjusts the quota of item exposures in a round-robin fashion based on user demands and past exposures, starting with a minimal initial quota. As shown in Algorithm~\ref{alg:rr}, RR efficiently adjusts the number of items recommended to users while ensuring no item is recommended more times than the dynamically calculated exposure limit. Starting from a minimal quota, the algorithm incrementally increases the allocation limit until it can satisfy the top recommendations for each user without violating the exposure constraints.
Then it allocates items to users in a greedy round-robin manner, ensuring that each user receives the top preferred items within the quota at each round, where the preferences ranking would be prefferably determined by the ranking scores output by the recommender, but if not available, could be detemined by any heuristic or random selection among the current recommended items.

\section{Additional Experimental Results}\label{app:additional_results}
\subsection{Comparison of moderation methods}\label{app:moderator_comparison}

\subsection{Pareto Frontier Plots}\label{app:pareto}

\subsection{Runtime}\label{app:runtime}

\section{Online Resources}
We have made our code and data available online at \url{https://anonymous.4open.science/r/RecSys_24_Submission_248}. %

\begin{table*}[]
    \begin{small}
    \caption{Comparison of moderation methods with \emph{inaccurate-CB} recommender \emph{without} user preference updating.}
    \begin{tabular}{lllllllllllll}
        \toprule
        & \multicolumn{3}{r}{S1. Balanced baseline} & \multicolumn{3}{r}{S2. Bi-polarized content} & \multicolumn{3}{r}{S3. Propaganda dominance} & \multicolumn{3}{r}{S4. Amplified right-stance} \\
        & CTR & JSD-O & JSD-G & CTR & JSD-O & JSD-G & CTR & JSD-O & JSD-G & CTR & JSD-O & JSD-G \\
        \midrule
        -          &                 0.628 &             0.047 &             0.159 &                    0.661 &             0.221 &             0.250 &                    0.628 &            0.333 &           0.321 &                      0.652 &             0.046 &             0.167 \\
        RR         &      0.553 (-12.1\%**) &   0.067** &   0.144ns &         0.565 (-14.5\%**) &   0.208** &   0.229** &          0.588 (-6.3\%**) &  0.331ns &  0.325ns &           0.574 (-12.1\%**) &    0.061* &   0.147* \\
        KC         &       0.582 (-7.4\%**) &   0.092** &    0.167ns &          0.605 (-8.5\%**) &  0.156** &  0.186** &          0.590 (-6.0\%**) &   0.360** &  0.342** &            0.594 (-9.0\%**) &   0.088** &   0.165ns \\
        RD         &       0.583 (-7.2\%**) &  0.026** &  0.118** &          0.607 (-8.1\%**) &   0.215ns &   0.230** &          0.610 (-2.8\%**) &   0.339ns &  0.325ns &            0.602 (-7.7\%**) &  0.033** &  0.125** \\
        SD         &       0.580 (-7.6\%**) &  0.041ns &  0.126** &          0.602 (-8.8\%**) &    0.228ns &   0.244ns &          0.612 (-2.5\%**) &   0.341ns &  0.327ns &            0.599 (-8.2\%**) &   0.053ns &  0.134** \\
        \bottomrule
    \end{tabular}    
    \end{small}
\end{table*}

\begin{table*}[]
    \begin{small}
    \caption{Comparison of moderation methods with \emph{random recommneder} recommender \emph{without} user preference updating.}
    \begin{tabular}{lllllllllllll}
        \toprule
            & \multicolumn{3}{r}{S1. Balanced baseline} & \multicolumn{3}{r}{S2. Bi-polarized content} & \multicolumn{3}{r}{S3. Propaganda dominance} & \multicolumn{3}{r}{S4. Amplified right-stance} \\
            & CTR & JSD-O & JSD-G & CTR & JSD-O & JSD-G & CTR & JSD-O & JSD-G & CTR & JSD-O & JSD-G \\
        \midrule
        -          &                 0.487 &             0.020 &             0.041 &                    0.483 &             0.206 &             0.210 &                    0.541 &            0.341 &            0.345 &                      0.495 &             0.020 &             0.040 \\
        RR         &       0.482 (-0.9\%ns) &  0.062** &  0.095** &          0.479 (-0.9\%ns) &    0.216ns &     0.225* &          0.535 (-1.2\%ns) &  0.316** &  0.322** &            0.494 (-0.4\%ns) &  0.056** &  0.092** \\
        KC         &       0.486 (-0.1\%ns) &  0.079** &  0.108** &           0.483 (0.1\%ns) &  0.136** &  0.161** &          0.536 (-0.9\%ns) &   0.342ns &   0.350ns &            0.492 (-0.8\%ns) &  0.079** &  0.107** \\
        RD         &       0.481 (-1.1\%ns) &    0.022ns &    0.043ns &          0.480 (-0.6\%ns) &    0.206ns &   0.209ns &          0.540 (-0.1\%ns) &  0.340ns &   0.345ns &            0.490 (-1.0\%ns) &   0.023ns &    0.044ns \\
        SD         &       0.469 (-3.7\%**) &  0.045** &   0.058** &           0.467 (-3.4\%*) &    0.208ns &    0.211ns &          0.540 (-0.2\%ns) &   0.341ns &   0.345ns &            0.477 (-3.8\%**) &  0.048** &   0.062** \\
        \bottomrule
    \end{tabular}    
    \end{small}
\end{table*}   

\begin{figure*}[]
    \centering
    \captionsetup{font=footnotesize}
    \begin{subfigure}{0.45\linewidth}
        \includegraphics[width=\linewidth]{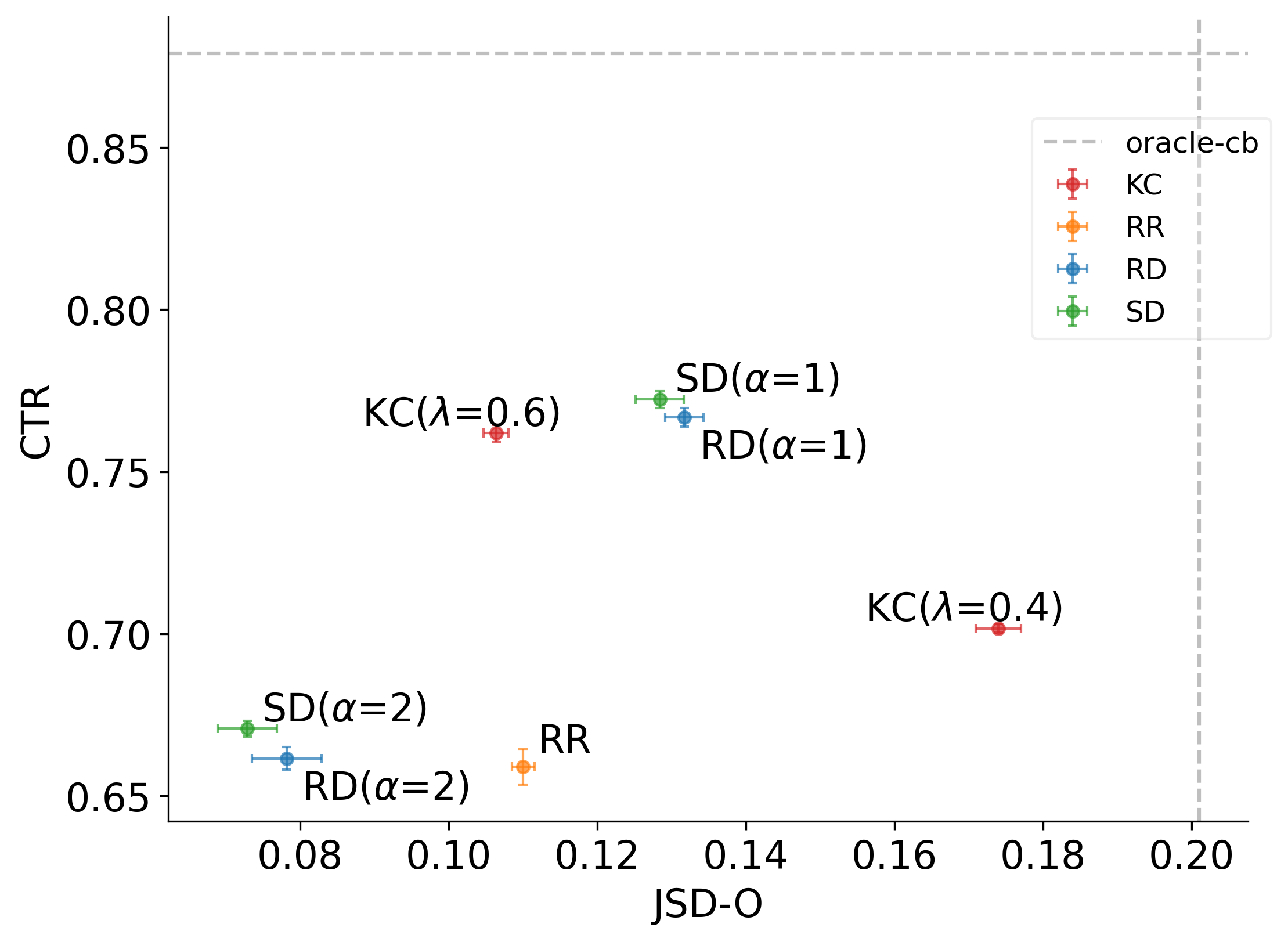}
        \caption{Scenario 1}
    \end{subfigure}
    \hfill
    \begin{subfigure}{0.45\linewidth}
        \includegraphics[width=\linewidth]{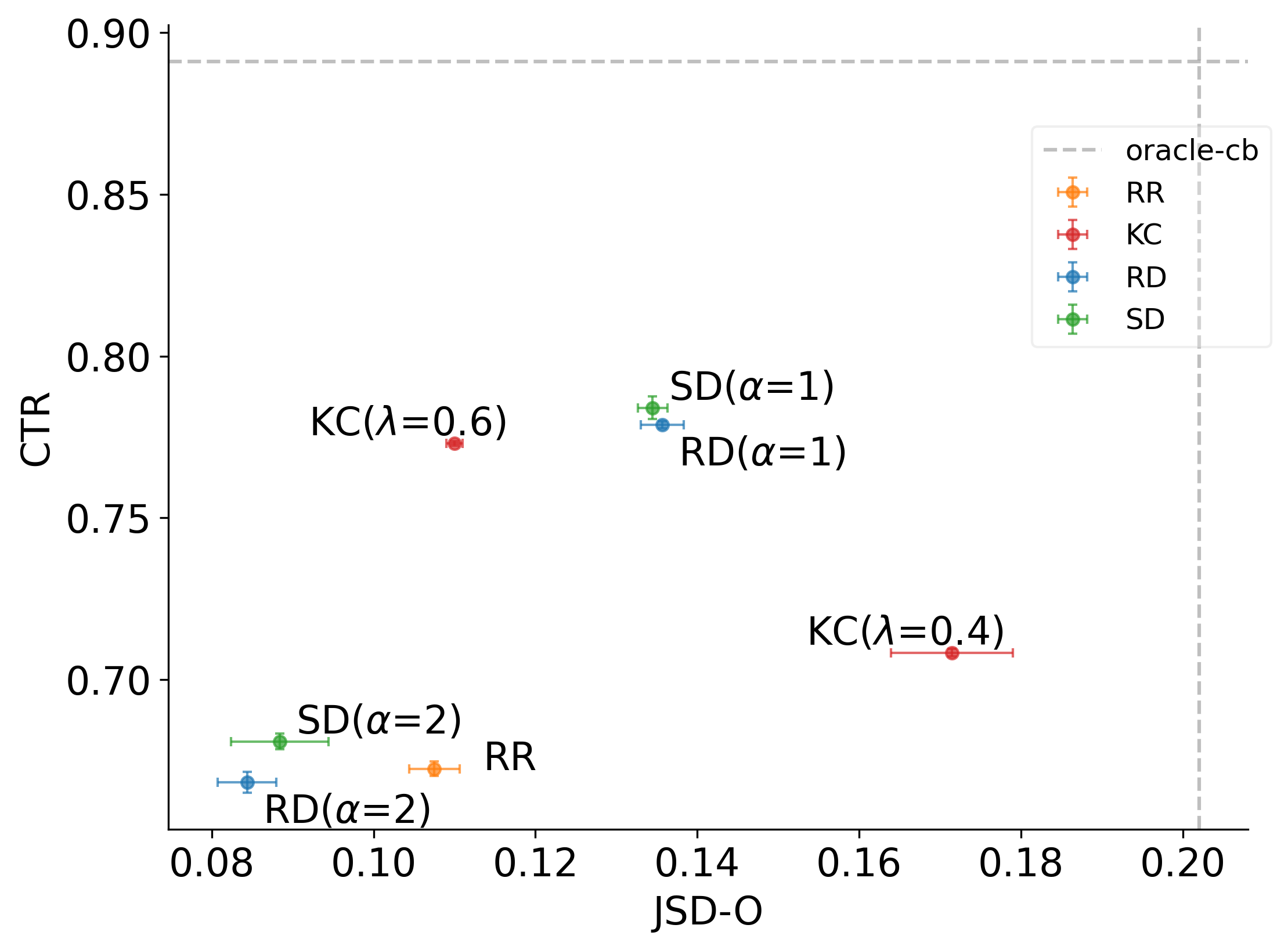}
        \caption{Scenario 4}
    \end{subfigure}
    \caption{Pareto frontiers of different moderation strategies with \emph{Oracle-CB} recommender and user preference \emph{updating}. Hyperparameters are bracketed, the meanings of which are explained in paragraph~\ref{sec:hyperparameters}. Markers are centered at the mean values over multiple runs, with errorbars roughly proportional to the standard deviations.}
    \label{fig:app_pareto_true}
\end{figure*}

\begin{figure*}[]
    \centering
    \captionsetup{font=footnotesize}
    \begin{subfigure}{0.45\linewidth}
        \includegraphics[width=\linewidth]{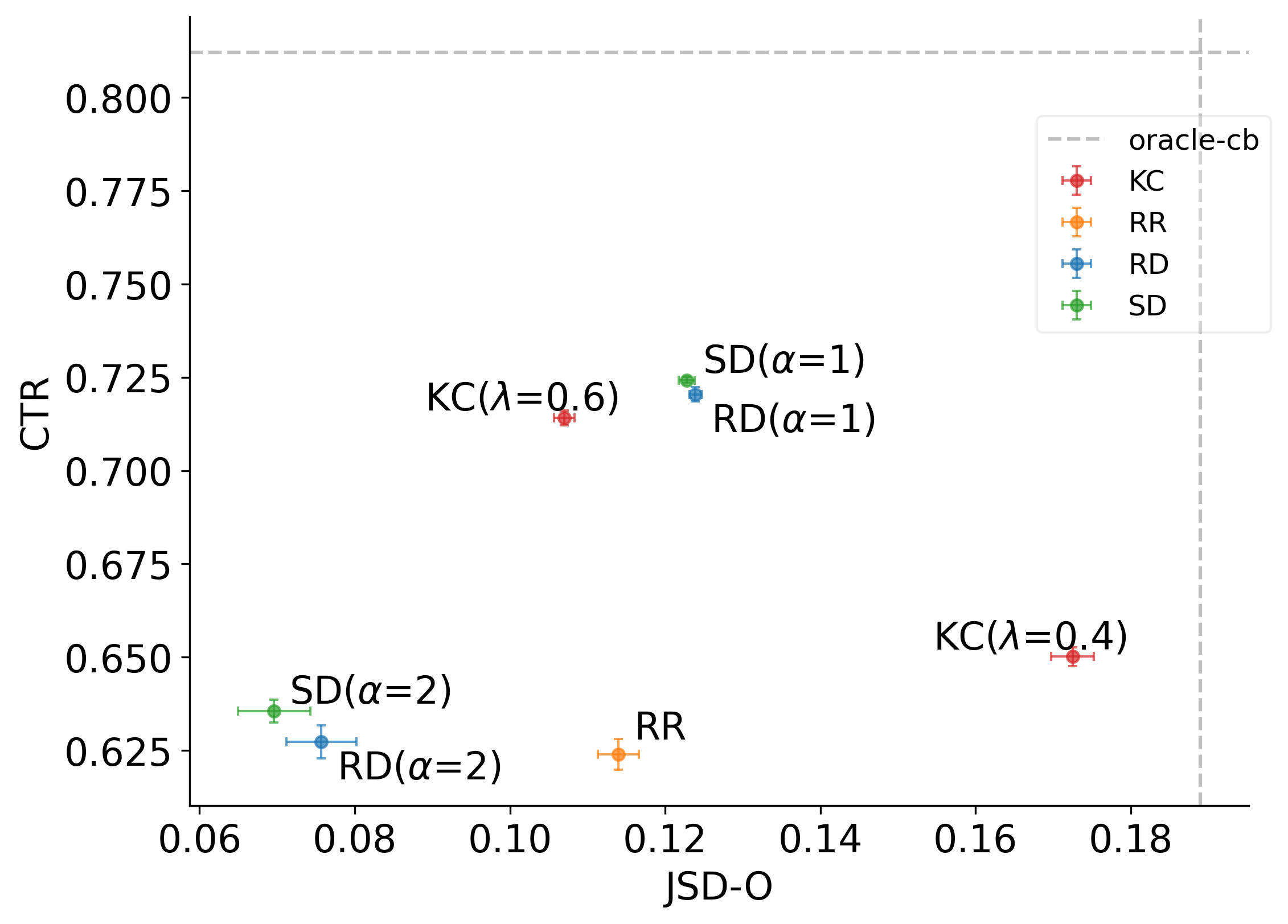}
        \caption{Scenario 1}
    \end{subfigure}
    \hfill
    \begin{subfigure}{0.45\linewidth}
        \includegraphics[width=\linewidth]{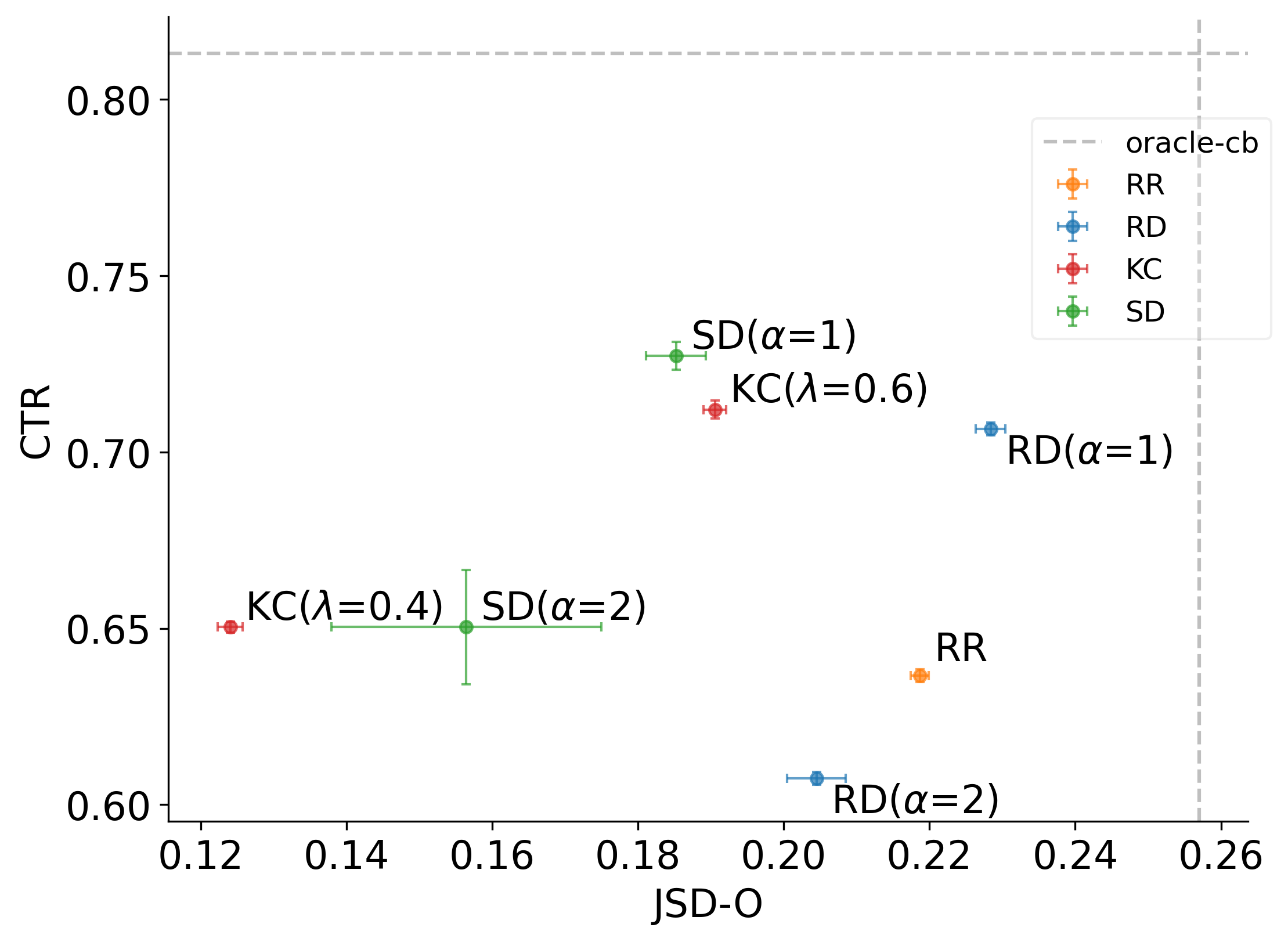}
        \caption{Scenario 2}
    \end{subfigure}
    \\
    \begin{subfigure}{0.45\linewidth}
        \includegraphics[width=\linewidth]{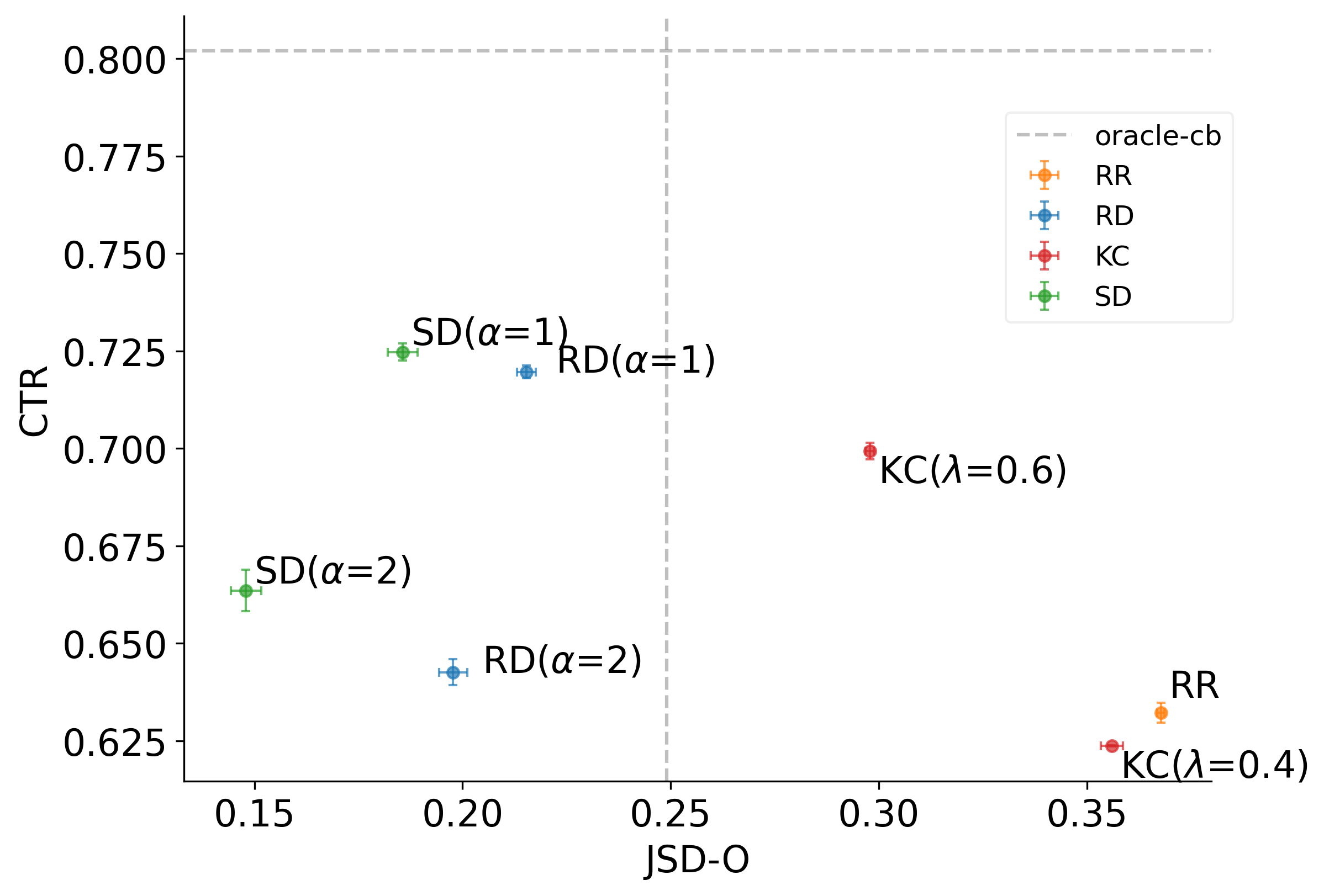}
        \caption{Scenario 3}
    \end{subfigure}
    \hfill
    \begin{subfigure}{0.45\linewidth}
        \includegraphics[width=\linewidth]{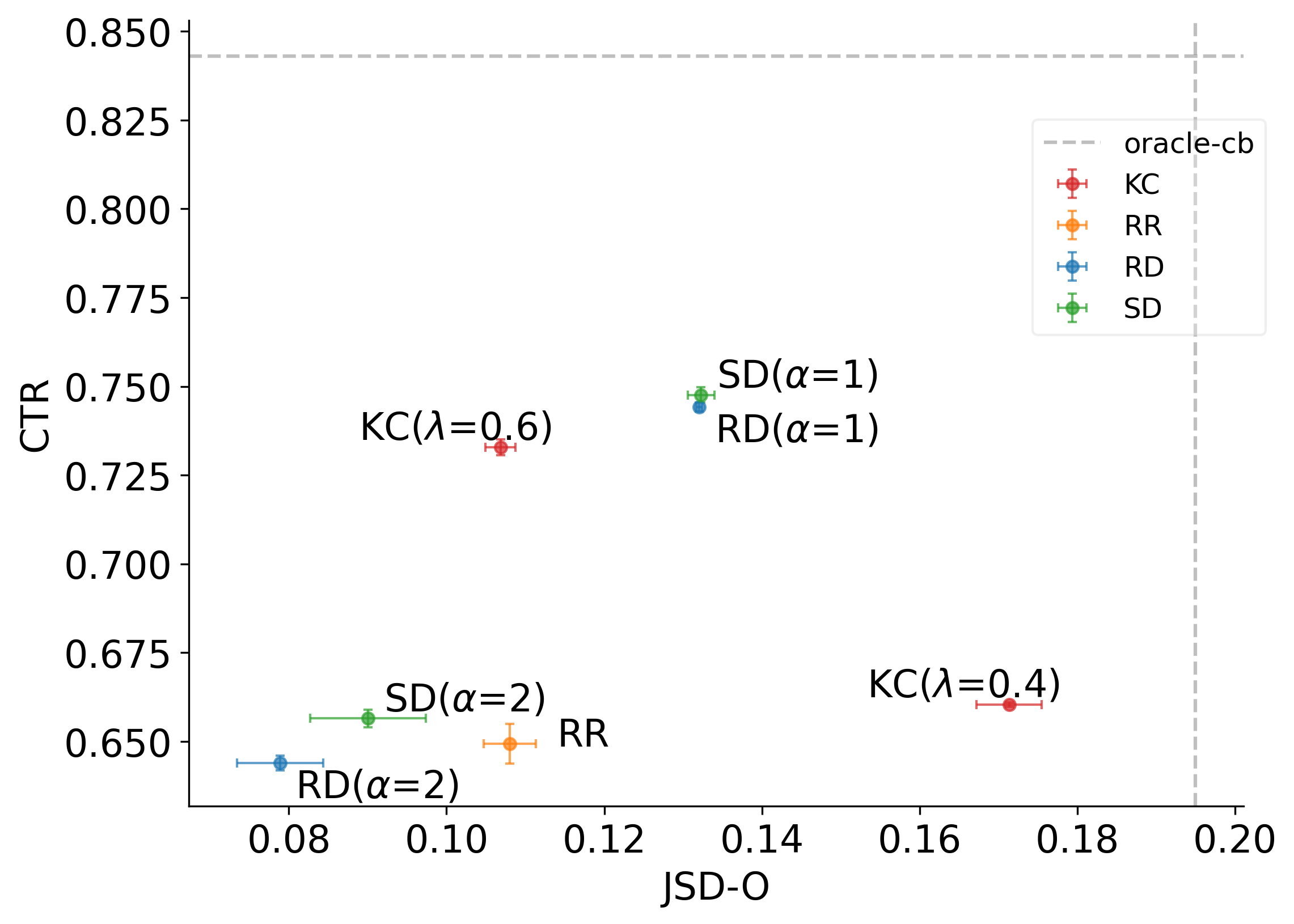}
        \caption{Scenario 4}
    \end{subfigure}
    \caption{Pareto frontiers of different moderation strategies with \emph{Oracle-CB} recommender and \emph{without} user preference updating.}
    \label{fig:app_pareto_false}
\end{figure*}

\begin{table*}[]
    \begin{small}
    \caption{Runtime comparison of moderation methods with different hyperparameters. The runtime per time step is measured in seconds. Each runtime is averaged over 30 random seeds per scenario, 4 scenario in total.}
    \begin{tabular}{llll}
        \toprule
        Moderator & Hyper-parameter &  Mean (seconds) &     STD \\         
        \midrule
        RR        &     - &              6.20 &   3.62 \\        
        KC       &   $\lambda=0.4$ &            181.01 &  21.09 \\
        KC       &   $\lambda=0.6$ &            138.65 &  59.47 \\
        RD     &     $\alpha=1$ &               0.74 &    0.24 \\
        RD     &     $\alpha=2$ &               0.72 &    0.37 \\
        SD     &     $\alpha=1$ &               0.60 &    0.29 \\
        SD     &     $\alpha=2$ &               0.68 &    0.57 \\
        \bottomrule
        \end{tabular}
    \end{small}
\end{table*}

\end{document}